\documentclass{article}

\usepackage{amsmath}
\usepackage{graphicx,psfrag,epsf} 
\usepackage{enumerate}
\usepackage{natbib}
\usepackage{url} 
\usepackage{rotating}
\usepackage{multirow}
\usepackage{amsfonts}
\usepackage{subfig}
\usepackage{epstopdf}
\usepackage{version}
\usepackage{amsthm}
\usepackage{gensymb}
\usepackage{multicol}
\usepackage{minitoc}
\usepackage[colorlinks=false, pdfborder={0 0 0}]{hyperref}
\usepackage{adjustbox}
\usepackage{float}
\usepackage{xcolor}

\author{Jose Ameijeiras--Alonso \quad Rosa M. Crujeiras \quad Alberto Rodr\'iguez--Casal}
\date{Department of Statistics, Mathematical Analysis and Optimization\\ \vspace{0.2cm} Universidade de Santiago de Compostela}						
\title{\pkg{multimode}: An \proglang{R} Package for Mode Assessment}

\newcommand{\pkg}[1]{{\normalfont\fontseries{b}\selectfont #1}}
\let\proglang=\textsf
\let\code=\texttt

\begin{document}

\maketitle

\begin{abstract}
In several applied fields, multimodality assessment is a crucial task as a previous exploratory tool or for determining the suitability of certain distributions. The goal of this paper is to present the utilities of the \proglang{R} package \pkg{multimode}, which collects different exploratory and testing nonparametric approaches for determining the number of modes and their estimated location. Specifically, some graphical tools, allowing for the identification of mode patterns, based on the kernel density estimation are provided (SiZer map, mode tree or mode forest). Several formal testing procedures for determining the number of modes are described in this paper and implemented in the \pkg{multimode} package, including methods based on the ideas of the critical bandwidth, the excess mass or using a combination of both. This package also includes a function for estimating the modes locations and different classical data examples that have been considered in mode testing literature.  
\end{abstract}

\noindent%

\section[On mode assessment]{A brief introduction on mode assessment}\label{intro}
Given a data sample from a random variable, determining the number of modes in the underlying density is a relevant question for supporting further decisions during the modelling approach. It is clear that unimodal distributions (such as the Gaussian density) may not be adequate for characterizing the behaviour of more complex data generating mechanisms in applied sciences. {Some examples requiring more complex distributions for reflecting the real number of modes can be found in many applied fields, such as astronomy, e.g. in the study of unimodal or multimodal patterns of the stars rotation periods for different temperatures \citep{mcquillan14}; business administration, e.g. when analysing the invested capital in crowdfunding campaigns \citep{colombo15}; forest science, e.g. in the analysis of the number of modes in the distribution of backscatter measurements (for unvegetated and dense forest areas), depending on the percentage of ground pixels \citep{santoro11}; genetics, e.g. for identifying which \textit{CpGs} (regions of DNA where a cytosine nucleotide is followed by a guanine nucleotide) present multimodal distributions \citep{Joubert16}; or psychology, where, for example, the study of the number of modes is crucial for detecting the presence of single or dual--process cognitive phenomena \citep{freeman13}; among others.} 

In principle, nonparametric density smoothers, such as the kernel density estimator introduced by \cite{Rosenblatt56} and \cite{Parzen62}, may overcome the problem of restricting the density estimation to a previously specified parametric family. Nevertheless, two important issues arise when performing density estimation via kernel (or any other) density smoothing methods. The first issue is that practitioners may be more comfortable interpreting and dealing with parametric models, since in many cases parameter estimates can be interpreted in terms of the data distribution given that they control some specific features. The second issue is that, even being satisfied with the nonparametric kernel density estimator output, since it provides an estimated version of the underlying distribution, there may be some doubts about the features highlighted by this curve estimator: are genuine from the distribution or are just due to sampling variability?

The previous concerns can be partially solved or answered by the identification of the (significant) modes in the kernel density estimator. Hence, as a previous step before fitting a parametric model, one should check how many distinguishable groups are there in the data distribution, being these groups identified by the modes of the density. This can be done by exploratory methods or by testing procedures, and in both cases, it should be also determined how much of the pattern observed in the density estimator is real, and how much is due to sampling artefacts. In addition, a very flexible and yet simple parametric approximation with several groups/modes can be carried out by fitting mixtures of normals \citep[a revision on this topic can be found in, for example,][]{McLachlan00}.

Quite a few contributions have been focused on solving the problem of identifying modes in a data distribution using nonparametric approaches, both from exploratory and testing perspectives. Regarding the exploratory approach, different proposals have been mainly focused on analysing the behaviour of the kernel density estimator along a range of different smoothing (bandwidth) parameters, where an \emph{expert eye} should try to identify \emph{persistent} patterns. The mode tree by \cite{MinSco93} and the mode forest \citep{Minetal98}, as well as the SIgnificant ZERo (SiZer) map by \cite{ChMar99} produce graphical displays where the change in the mode pattern of the density estimator can be clearly seen along different bandwidth values.

The aforementioned exploratory tools, although providing a complete analysis of the density estimate from a scale--space perspective \citep[see][]{ChMar99}, require a decision on the number of modes to be taken after examining a graphical output. Therefore, conclusions cannot be directly obtained by applying an automatic procedure which indicates how many of the modes observed in the previous representations are really significant. However, this question can be answered by a hypothesis test: $H_0:\;j=k$ vs. $H:\; j>k$, denoting by $j$ the real number of modes in the density and being $k$ a positive integer (so $k=1$ is a unimodality test). This testing problem has been solved designing test statistics which are based on the critical bandwidth \citep{Silverman81,HallYork01,FisMar01} and/or the excess mass \citep{Hartigan85,MulSaw91,ChengHall98,Ameijeirasml}. These procedures will be briefly described in the paper, along with the previous exploratory methods.

{Some of the parametric and nonparametric tools for exploring the number of modes on a data distribution are already implemented in other packages in the \textit{CRAN} repository of \proglang{R} \citep{R18}. A brief summary of the capabilities of some packages are provided below. The aim of the \proglang{R} package presented in this paper, \pkg{multimode} \citep{Ameijeiraspkg}, is to provide an easy--to--use toolbox with different nonparametric methods for assessing multimodality in real distributions. The methods included in the package facilitate both the exploratory and inferential analysis.}

\begin{itemize}
\item \pkg{diptest} \citep{Maechler13}: This package is focused in the \textit{dip} test of \cite{Hartigan85}, which allows for testing unimodality against multimodality.
\item \pkg{feature} \citep{Duong15}: Based on the SiZer map, this package provides some exploratory tools for detecting where the smoothed curve is significantly increasing or decreasing for the 1--dimensional case \citep[with similar ideas to][]{ChMar99}, 2--dimensional \citep{godtliebsen02} and also for the 3 and 4--dimensional cases \citep{duong08}. 
\item \pkg{mixtools} \citep{Benaglia09}: This package includes different parametric methods based on finite mixture models. Among other functionalities, it allows for testing or exploring the number of components on finite mixture models \citep[][Ch. 6]{McLachlan00}. In particular, it computes different information criteria (\code{multmixmodel.sel}, \code{repnormmixmodel.sel} and \code{regmixmodel.sel}) and it performs a parametric bootstrap for testing a $m$--component versus a $(m+1)$--component fit (\code{boot.comp}) for mixtures of multinomials, multivariate normals and some kinds of regression models.
\item \pkg{modeest} \citep{Poncet12}: When knowing that the underlying distribution of the data is unimodal, this package provides different parametric and nonparametric methods for estimating the mode location.
\item \pkg{modehunt} \citep{Rufibach15}: This package implements some nonparametric methods that do not employ the kernel density estimation and, therefore, do not depend on the bandwidth parameter \citep{dumbgen2008,rufibach2010}. Based on the ordered sample, the methods provide open intervals, with endpoints at data points, for which the density function $f$ is significantly increasing or decreasing.
\item \pkg{NPCirc} \citep{Oliveira14b}: Among other functionalities, this package, with functions \code{circsizer.density} and \code{circsizer.regression}, extends the SiZer map to the context of circular data, i.e., samples that can be represented as points on the circumference of a unit circle \citep{oliveira14c}.  
\end{itemize}

There are different combinations of views and goals that must be considered when proceeding with multimodality assessment. First, a parametric or a non parametric approach can be used. Then, it may be enough with an exploratory tool for determining the number of modes or maybe a formal testing procedure could be required. Finally, it may be crucial also to determine the modes locations.  

{First, if the parametric approach is chosen, package \pkg{mixtools} provides different techniques for determining the number of modes in this context. Following a nonparametric perspective, available methods in \proglang{R} are based in the ordered sample (package \pkg{modehunt}) or in density smoothing approaches. }

{As observed in the previous analysis of the different \proglang{R} packages, just a few techniques are available for identifying the number of modes using the kernel density estimation. In particular, if the exploratory way is chosen, package \pkg{feature} provides some graphical methods (based on the SiZer map) and package \pkg{diptest} the testing approach of \cite{Hartigan85}. The objective of the functions in \pkg{multimode} is complementing other implementations on nonparametric multimodality analysis. When referring to other statistical software languages, up to the authors' knowledge, besides the aforementioned non--parametric proposals, just the \cite{Silverman81} testing approach was already available \citep[see, e.g. \code{silvtest} function in \proglang{Stata};][]{Salgado98}.}

When focusing on graphical methods, apart from the SiZer, \pkg{multimode} provides other exploratory methods, such as the mode tree and the mode forest. Referring to the SiZer map, the main difference with function \code{SiZer} of \pkg{feature} is the way of calculating the confidence intervals for the derivative of the kernel density estimation. While in \pkg{feature}, its own approximation is performed, the four proposed methods by \cite{ChMar99} (based on normality and bootstrap techniques) for calculating where the smoothed curve is significantly increasing/decreasing are provided in \pkg{multimode}. In Figure~\ref{figapmm2}, the differences between both packages can be observed (\code{SiZer} of \pkg{feature} in panel g, \code{sizer} of \pkg{multimode} in panels e, f, h and i). Note that, for representing the bandwidth values, although \pkg{feature} uses a base $e$ instead of the base 10 logarithm \citep[the last one suggested by][]{ChMar99}, for comparative purposes, in this case, both are given in $\log_{10}$ scale. The SiZer maps are represented using a sample including the thickness of stamps (introduced in Section~\ref{data_description}) where at least two modes are expected \citep[see][]{IzenSom88}. Modes in SiZer can be detected by blue--red patterns (see Section~\ref{exploratory_tools}). Hence, the SiZer obtained from the \pkg{feature} package (and, also, using the Gaussian approximations in \pkg{multimode}) detects at most just one mode, while more than one mode can be observed in the SiZer maps obtained from \pkg{multimode} with bootstrap methods (see Section~\ref{exploring_data}). 

Apart from the unimodality test of \cite{Hartigan85} (already implemented in \pkg{diptest} package), \pkg{multimode} includes several proposal for testing the number of modes. Since the dip test presents an extremely conservative behaviour \citep[see][]{Ameijeirasml}, the objective here is including other proposals and provide a way of testing a general number of modes. 

{Finally, when the objective is to estimate the modes locations, the aforementioned graphical tools already provide a way of exploring their locations (depending on the bandwidth parameter). In the situation of having a unimodal distribution, package \pkg{modeest} includes some (parametric and nonparametric) tools for estimating the mode location. Also, when the (general) number of modes is known, package \pkg{multimode} also provides a (nonparametric) way of estimating the modes (and antimodes) locations.}

{With the objective of presenting how to tackle the problem of identifying the number and locations of modes and showing the capabilities of the \pkg{multimode} package, this paper is organized as follows:} in Section 2, some background on both exploratory and testing methods for assessing multimodality will be provided. Initially, the kernel density estimator will be briefly introduced, as it is the key tool for the exploratory and testing methods to be presented. In this section an overview of different graphical tools (namely, the mode tree, the mode forest and the SiZer map) will be provided. Also, different procedures for testing the number of modes are described, including those ones using the critical bandwidth or the excess mass. In Section 3, the reader will find a guided tour across \pkg{multimode}, illustrating its use with a real data example. Finally, some discussion will be provided in Section 4, commenting also on the possible extensions of the package.

\section[Exploratory and testing methods]{Exploratory and testing methods for assessing multimodality}
This section provides a brief background on the design of the different (exploratory and testing) tools included in \pkg{multimode}. A key element in the foundations of the different proposals is the kernel density estimator. Given a random sample $(X_1,\ldots,X_n)$ from a random variable $X$ with (unknown) density $f$, the kernel density estimator for a fixed $x\in \mathbb{R}$ is defined as:
\begin{equation}
\hat{f_h}(x)=\frac{1}{nh}\sum_{i=1}^nK\left(\frac{x-X_i}{h}\right),
\label{kernel_estimator}
\end{equation}
where $K$ is the kernel function (usually a symmetric and unimodal density) and $h>0$ is the smoothing parameter or bandwidth. This parameter controls the smoothness of the estimator in the sense that large (small) values of $h$ provide oversmoothed (undersmoothed) curves. For the particular case of a Gaussian kernel, and focusing on the modes exhibited by $\hat{f_h}$, it should be noted that the number of modes is monotone in $h$ \citep{Silverman81}. This feature is essential to guarantee the validity of the different proposals.

\subsection[Exploratory tools]{Exploratory tools}\label{exploratory_tools}

\begin{figure}
\begin{multicols}{3}
\centering
\includegraphics[width=1\linewidth]{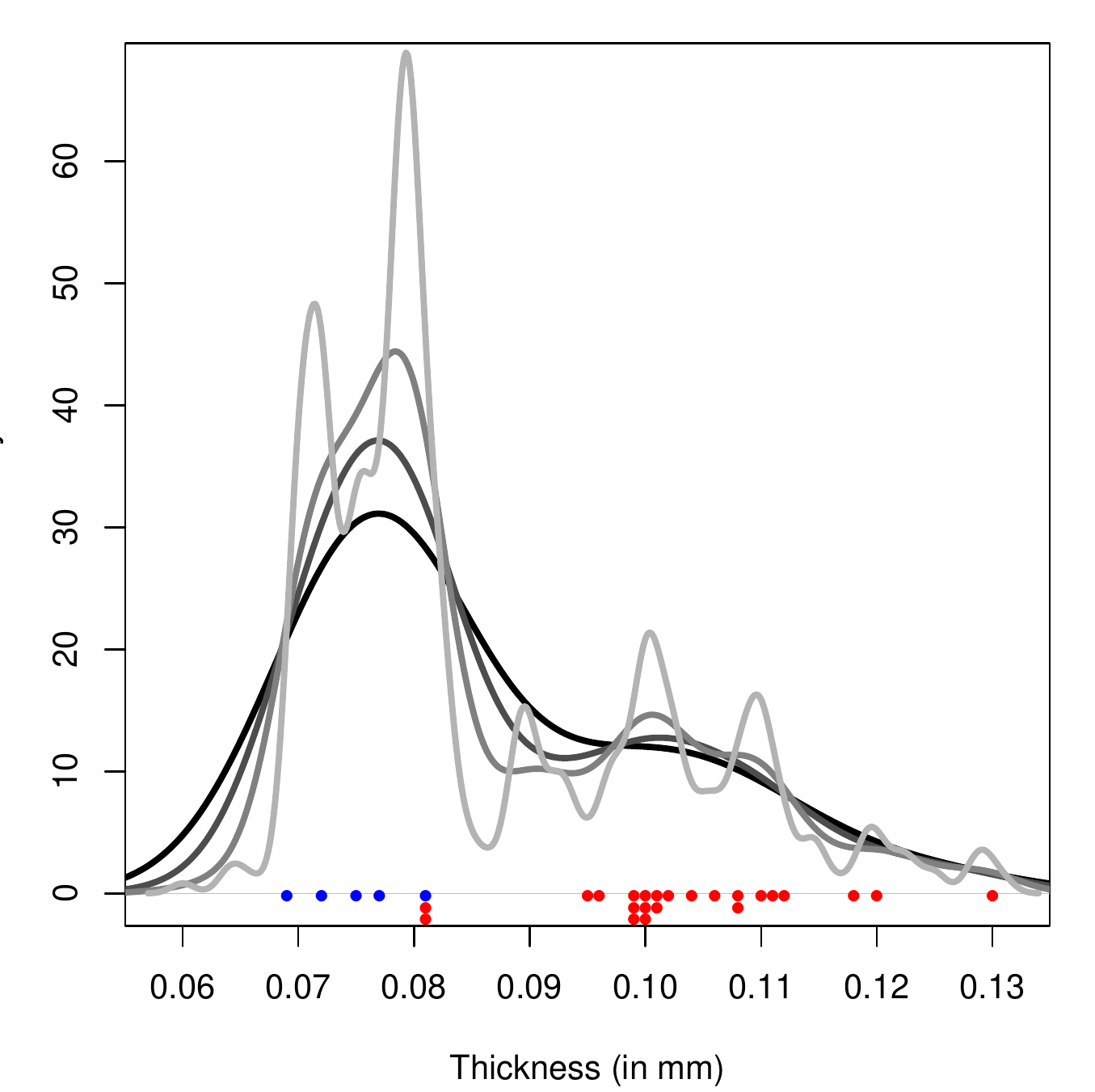}\\
(a)
\includegraphics[width=1\linewidth]{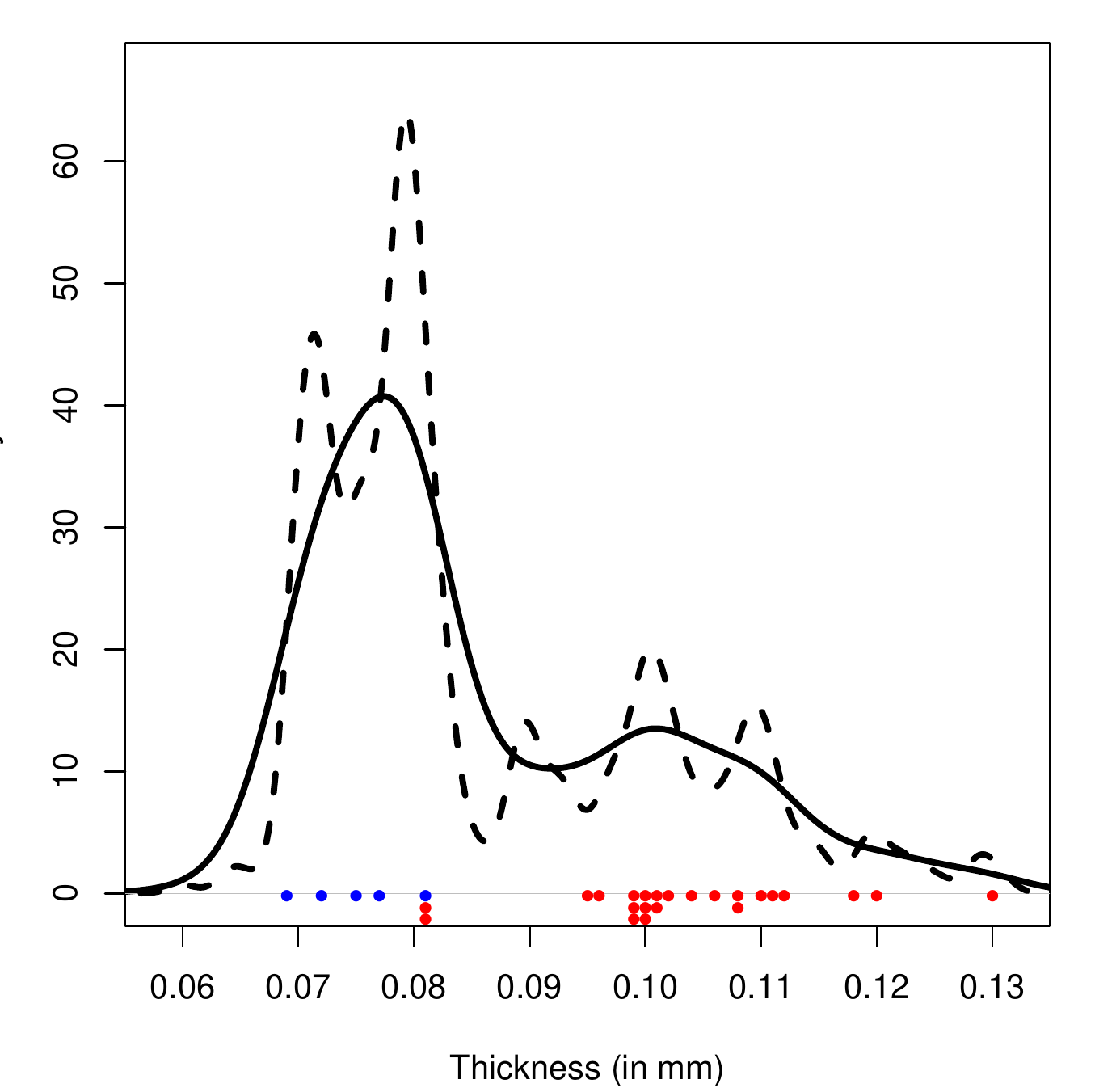}\\
(d)
\includegraphics[width=1\linewidth]{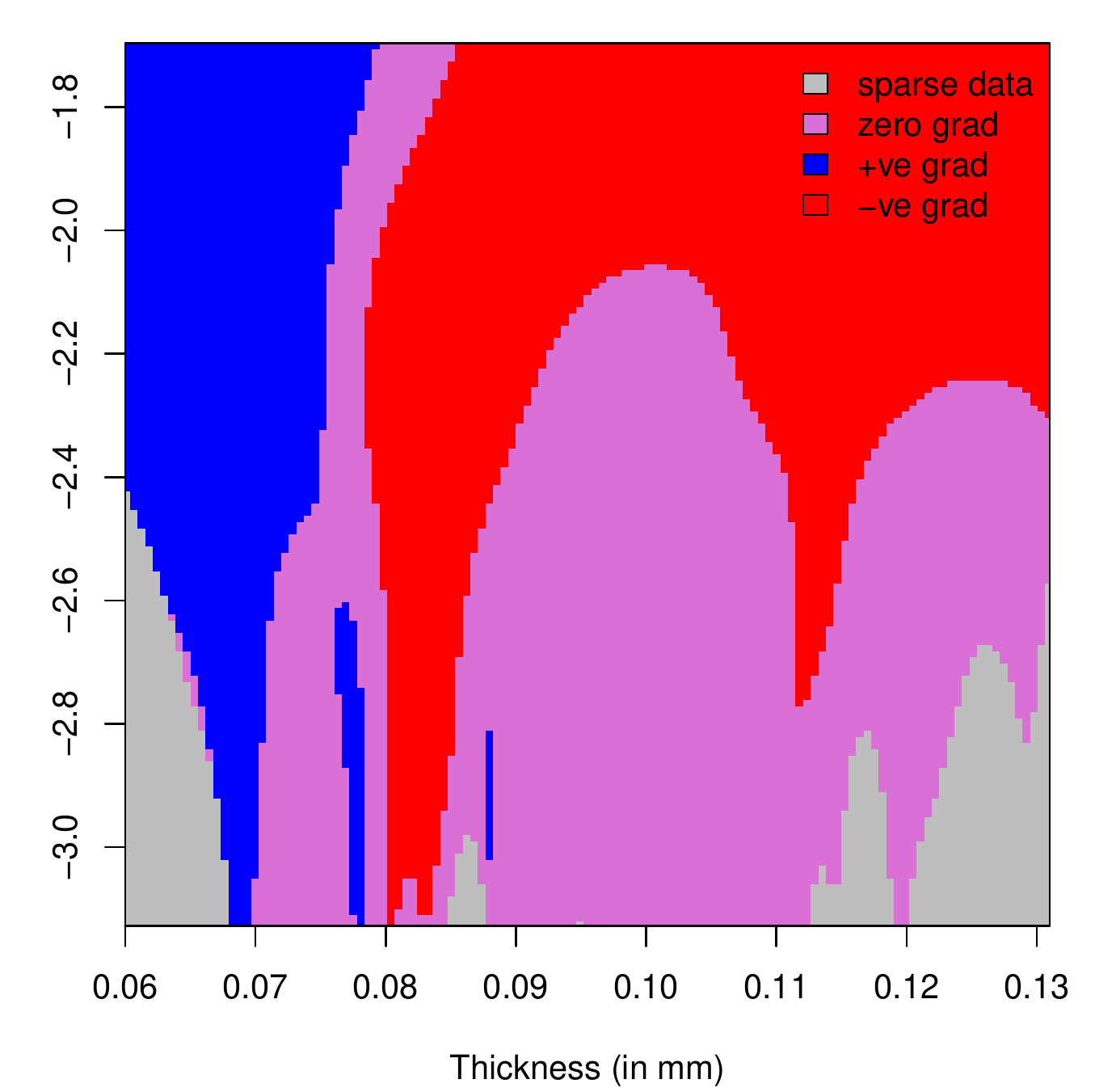}
(g)\\
\includegraphics[width=1\linewidth]{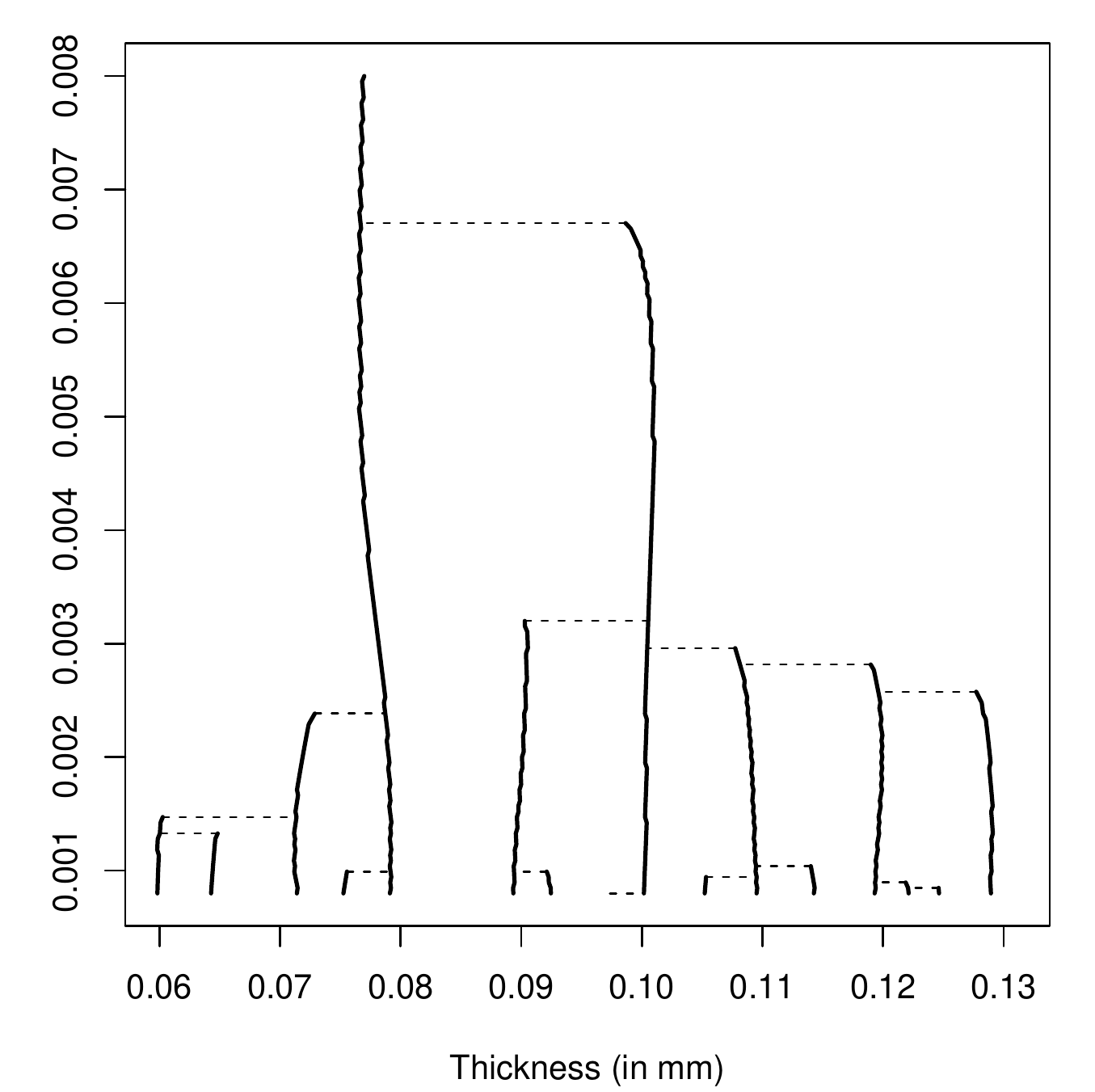}\\
(b)
\includegraphics[width=1\linewidth]{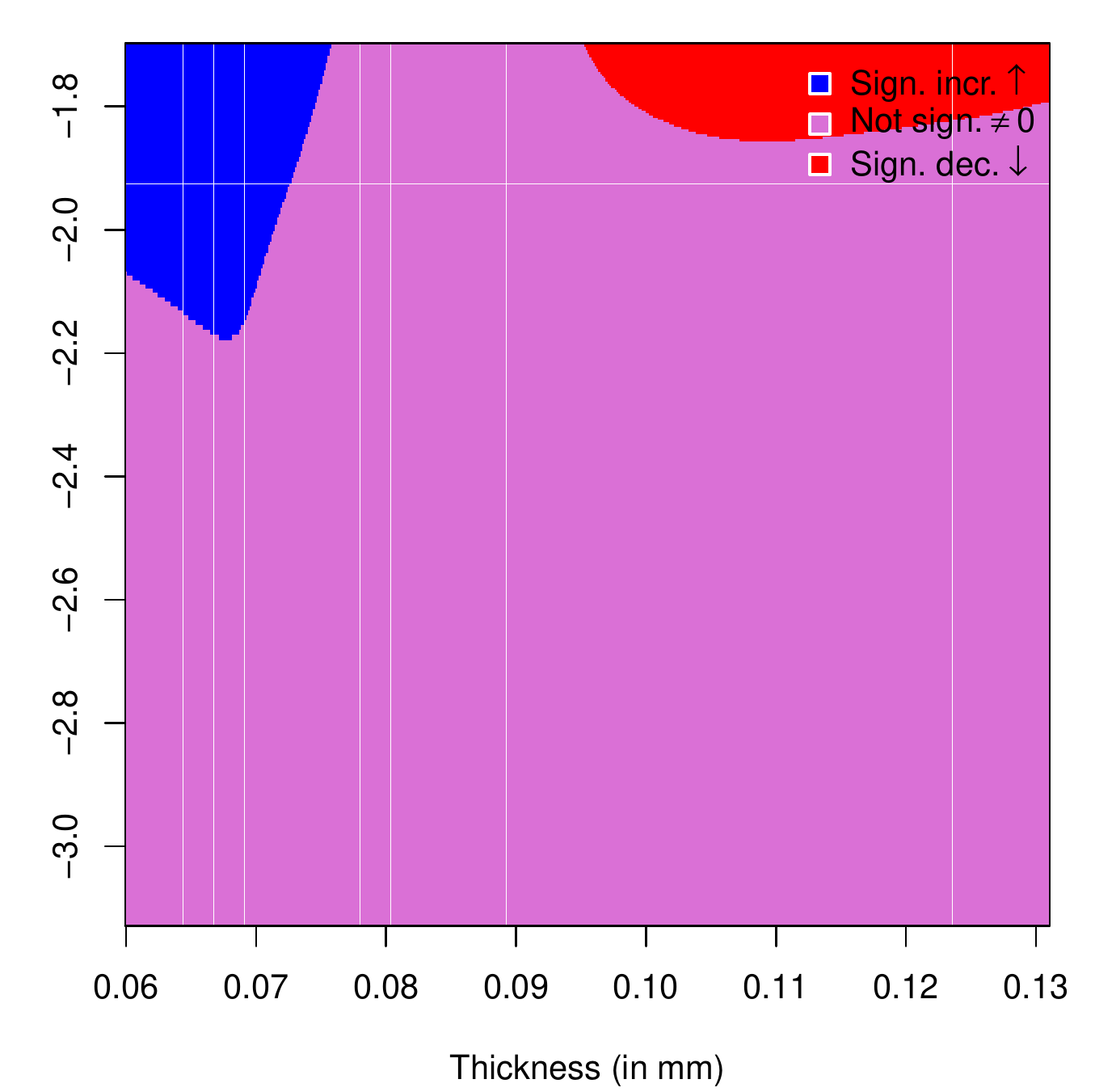}\\
(e)
\includegraphics[width=1\linewidth]{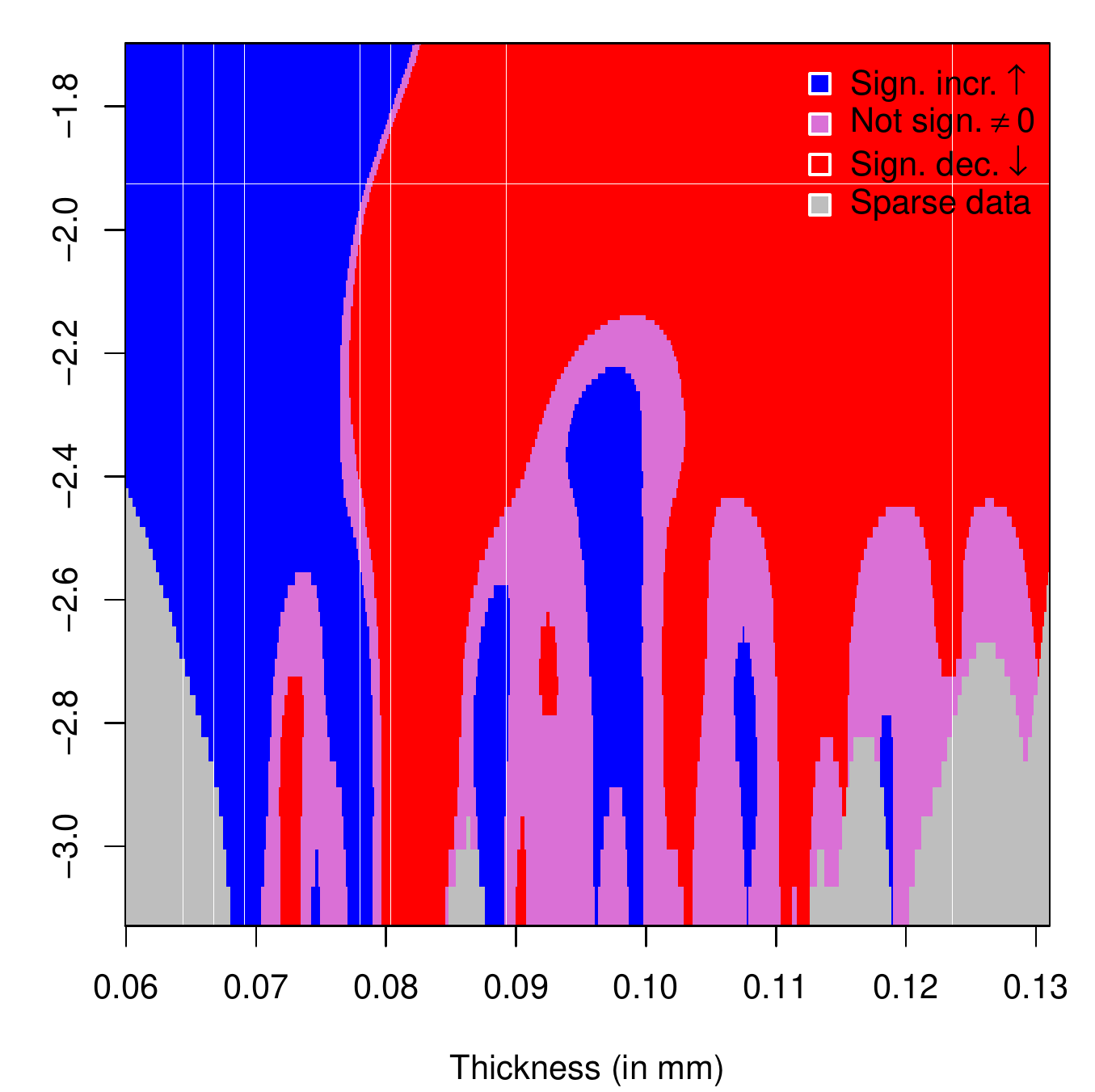}
(h)\\
\includegraphics[width=1\linewidth]{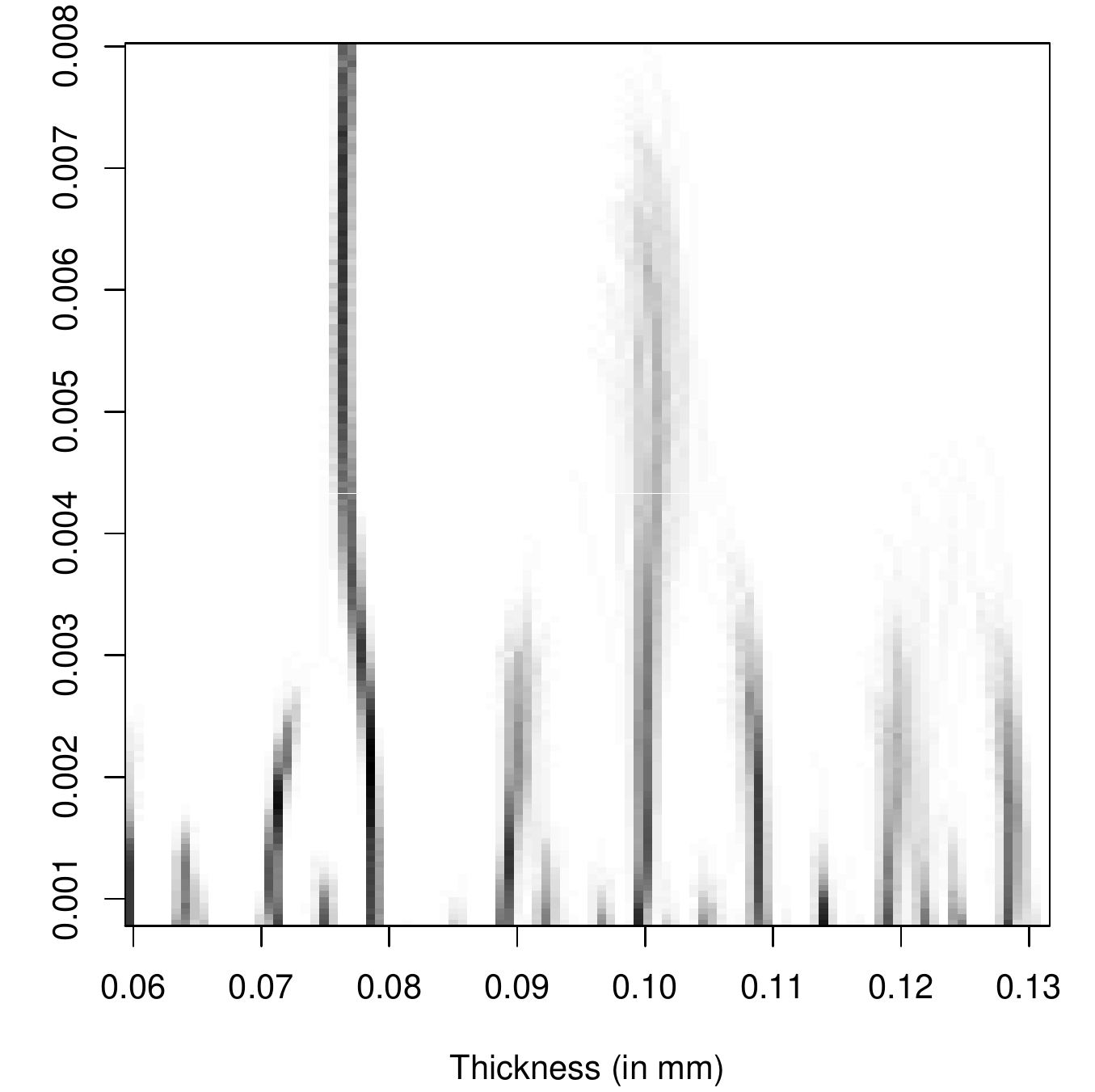}\\
(c)
\includegraphics[width=1\linewidth]{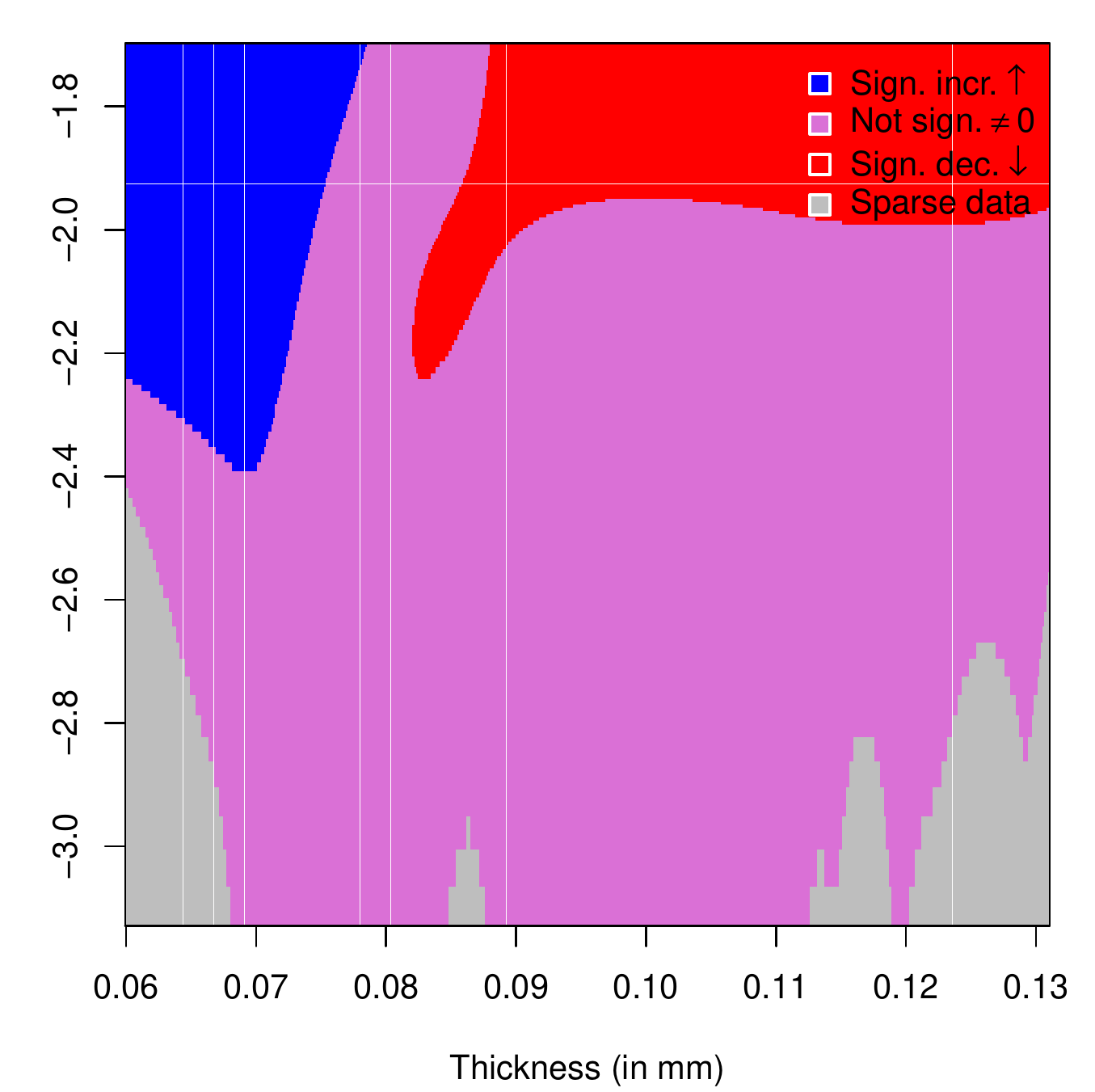}\\
(f)
\includegraphics[width=1\linewidth]{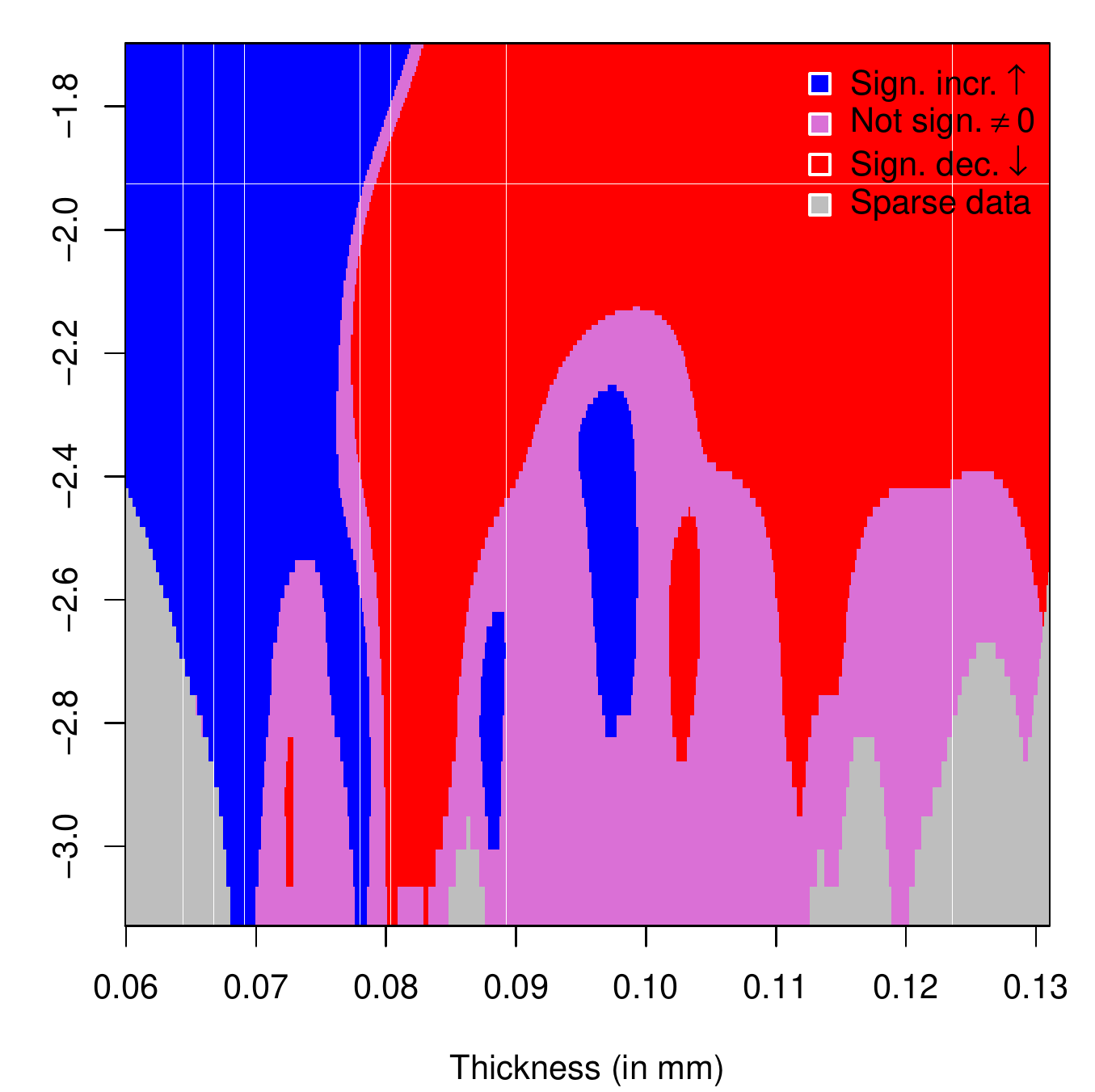}
(i)\\
\end{multicols}
 \caption{Exploratory analisys for a sample of 485 stamps (1872 Hidalgo Issue of Mexico). In (a) and (d), kernel density estimators with Gaussian kernel and different bandwidths; the points represent stamps watermarked with \textit{LA+-F} (blue) and \textit{Papel sellado} (red). In (a), from dark to light grey, $h$ values: 0.007, 0.005, 0.003 and 0.001. In (d), $h=0.0039$ (rule of thumb -continuous line-) and $h=0.0012$ \citep[plug--in rule -dashed line-, see][Ch. 3]{wandjones}. Mode tree (b) and mode forest (c) between the bandwidths $8\cdot10^{-3}$ and $8\cdot10^{-4}$. For each $h$, the estimated modes locations are identified by continuous lines in (b) and dark grey pixels in (c). The horizontal discontinuous lines (b) indicate how each mode splits. Panels (e)--(i): SiZer maps between $\log_{10}(h)=-1.7$ ($h=0.02$) and $\log_{10}(h)=-3.1$ ($h=8\cdot10^{-4}$); given a value of $\log_{10}(h)$, modes can be detected by blue--red patterns. Obtained from \pkg{feature} package (g), using Gaussian, $q_1$ (e) and $q_2$ (f), and bootstrap, $q_3$ (h) and $q_4$ (i), quantiles.}
 \label{figapmm2}
\end{figure}

Since the number of modes in $\hat{f_h}$ is a monotone decreasing function of $h$, when the Gaussian kernel is used, a simple exploratory solution, for determining the number of modes, is representing this density estimation for different values of $h$ (see Figure~\ref{figapmm2}, panel a). In fact, this is the idea underlying some graphical tools, such as the mode tree and the mode forest, where an example of both representations is provided in Figure~\ref{figapmm2} (panels b and c).

In the mode tree, \cite{MinSco93} created a tree diagram (similar to the dendrogram) representing, with continuous vertical lines, the modes locations (primary axis) of $\hat{f_h}$ for different bandwidth parameters $h$ (secondary axis). In addition, it represents, with horizontal dashed lines, how each mode splits into more modes as the bandwidth decreases (from top to bottom), showing the relationship between the new modes and the original modes from which they split.  

As pointed out by \cite{Minetal98}, the problem of the mode tree is the strong dependence on the available sample. That is the reason why the mode forest is constructed by computing the position of the estimated modes from different mode trees obtained from sampling with replacement the original sample. In order to facilitate the visualization of this exploratory method, the graphical window is divided in different (previously chosen) location--bandwidth (horizontal--vertical axis) pixels. Then, this tool represents the number of times that an estimated mode falls in each (location--bandwidth) pixel shading it proportionally to counts (large counts corresponding to darker pixels and low counts to lighter ones). Then, in the mode forest, modes are identified by dark grey regions.

A problem of the mode tree and the mode forest is that they do not identify which modes are artificially created by atypical data points. An exploratory tool that avoids this issue is the SiZer proposed by \cite{ChMar99} and whose representation can be observed in Figure~\ref{figapmm2} (panels e, f, h and i). SiZer identifies the significant features of the density, by analysing the behaviour of the derivative of the kernel density estimation. For a given location (horizontal axis) and using a specified bandwidth parameter (vertical axis), the SiZer map represents where the smoothed curve is significantly increasing (blue colour), decreasing (red) or not significantly different from zero (orchid, a light tone of purple). Thus, for a given bandwidth, a significantly increasing region followed by a significantly decreasing region (blue--red pattern) indicates where a significant peak is present.

For determining the behaviour of the smoothed curve, fixing a location $x$ and a bandwidth $h$, the confidence limits of $\hat{f}_{h}'(x)$ are of the form $\mbox{CI}^{\pm}(x,h)=\hat{f}_{h}'(x) \pm \mbox{quantile} (\alpha) \cdot \widehat{\mbox{sd}} (\hat{f}_{h}'(x))$, where $\widehat{\mbox{sd}}$ is the estimated standard deviation and $\alpha$ is the significance level. The estimation of the variance of $\hat{f}_{h}'(x)$ is obtained in the following way

\begin{equation}\label{varsizer}
\widehat{\mbox{var}} \left(\hat{f}_{h}'(x) \right)= \frac{1}{n h^4} S^2 \left( K'\left(\frac{x-X_1}{h}\right), ..., K'\left(\frac{x-X_n}{h}\right) \right),
\end{equation}

where $S^2$ in (\ref{varsizer}) denotes the sample variance. In order to calculate the quantiles, \cite{ChMar99} proposed four approximations: two based on Gaussian methods and two based on bootstrap techniques. The first proposal is based on pointwise Gaussian quantiles ($q_1$; Figure~\ref{figapmm2}, panel e), where quantiles are calculated as $q_1(\alpha)=\Phi^{-1}\left( 1- {\alpha}/{2} \right)$, being $\Phi^{-1}$ the normal quantile function. The second method provides approximate Gaussian quantiles simultaneous over $x$ ($q_2$; Figure~\ref{figapmm2}, panel f) and they are defined as $q_2(\alpha;h)=\Phi^{-1}\left(  {1+(1-\alpha)^{1/m(h)}}/{2} \right)$. For each bandwidth, $m(h)$ are obtained from the Effective Sample Size \citep[\textit{ESS}, see][]{ChMar99} in the following way

\begin{equation}\label{ESSsizer}
m(h)=\frac{n}{\overline{\mbox{ESS}}(x,h)}, \quad \mbox{being } \mbox{ESS}(x,h)=\frac{\sum_{i=1}^nK\left(\frac{x-X_i}{h}\right)}{K(0)}
\end{equation} 
and $\overline{\mbox{ESS}}(x,h)$ the average mean over $x$ of the values of $\mbox{ESS}(x,h)$. Small values of ESS provide an indicative of areas with too sparse data for meaningful inference. For that reason, in the methods employing the ESS ($q_2$, $q_3$ and $q_4$), the significant features are just represented in the regions satisfying $x\in D_h=\{x: \mbox{ESS}(x,h) \geq n_0\}$ (remaining regions are marked with grey colour; see Figure~\ref{figapmm2}, panels f, h and i). Then, the parameter $n_0$ \citep[where][proposed to use $n_0=5$]{ChMar99} plays a fundamental role for removing the spurious modes created by atypical data points. 

The two bootstrap quantiles are calculated from the following values

\begin{equation}\label{Zxhbsizer}
Z(x,h)^{*b}=\frac{\hat{f}_{h}'(x)^{*b} - \hat{f}_{h}'(x)}{\widehat{\mbox{sd}} (\hat{f}_{h}'(x))}, \mbox{ with } b=1,...,B,
\end{equation}
where each $\hat{f}_{h}'(x)^{*b}$ is calculated from a random sample generated drawn with replacement from the original sample. The third approach is a bootstrap quantile simultaneous over $x$, $q_3(\alpha;h)$ (Figure~\ref{figapmm2}, panel h), and it is calculated with the empirical quantile $(1-\alpha/2)$ of the $B$ values ${\max}_{x \in D_h}  |Z(x,h)^{*b}|$; with $b=1,...,B$. Finally, the fourth approach, also calculated from the quantities defined in (\ref{Zxhbsizer}), is the bootstrap quantile simultaneous over $x$ and $h$, $q_4(\alpha)$ (Figure~\ref{figapmm2}, panel i), and it is defined as the empirical quantile $(1-\alpha/2)$ of the $B$ values ${\max}_h {\max}_{x \in D_h} |Z(x,h)^*|$; with $b=1,...,B$.

\subsection[Testing procedures]{Testing procedures}\label{testing_procedures}
Consider the testing problem presented in the Introduction. That is, given a sample $X_1,\ldots,X_n$ from a random variable $X$ with unknown density $f$ with $j$ modes, and given a positive integer $k$, the goal is to test $H_0:\;j=k$ vs. $H_a;\;j>k$. The testing methods, briefly described in this section and included in \pkg{multimode}, make use of one or both of the following concepts: the critical bandwidth and the excess mass.

\subsection{Using a critical bandwidth}
\label{background:critical}
The critical bandwidth for a fixed $k$ was defined by \cite{Silverman81} as the smallest bandwidth such that the kernel density estimator in (\ref{kernel_estimator}) has at most $k$ modes:
\begin{equation}
h_k=\inf\{h :\hat{f_h} \mbox{ has at most } k \mbox{ modes}\}.
\label{critical_bandwidth}
\end{equation}
This value can be used as a test statistic, as long as (\ref{kernel_estimator}) is constructed with a Gaussian kernel, as proposed by \cite{Silverman81}: $H_0$ is rejected for large values of $h_k$. For calibrating $h_k$, a bootstrap algorithm is employed, where the resamples $Y^{*b}_i=(1+h_k^2/\hat{\sigma}^2)^{-1/2} X^{*b}_i$ (with $i\in\{1,\ldots,n\}$, being $n$ the sample size) are calculated from $B$ bootstrap samples $X^{*b}_i$ generated from $\hat{f}_{h_k}$, being $\hat{\sigma}^2$ the sample variance and with $b\in\{1,\ldots,B\}$. \cite{HallYork01} proved that this bootstrap algorithm is not consistent and the authors suggested a correction for the unimodality test (for $k=1$), when $f$ has a bounded support or when the mode is located in a given closed interval $I$, defining the critical bandwidth as:
\begin{equation}
h_{\tiny{\mbox{HY}}}=\inf\{h :\hat{f_h} \mbox{ has exactly one mode in } I\}.
\label{bw_hy}
\end{equation}
The authors also proposed using $h_{\tiny{\mbox{HY}}}$ as a test statistic and designed a bootstrap algorithm in this simplified scenario. However, the critical bandwidths for the bootstrap samples $h_{\tiny{\mbox{HY}}}^{*}$, calculated from $X^{*}$, are smaller than $h_{\tiny{\mbox{HY}}}$, so for an $\alpha$--level test, a correction factor $\lambda_{\alpha}$ to empirically approximate the p--value $\mathbb{P}(h_{\tiny{\mbox{HY}}}^* \leq \lambda_\alpha h_{\tiny{\mbox{HY}}} | \mathcal{X}) \geq 1- \alpha$ must be considered. Two different methods were suggested for computing this $\lambda_{\alpha}$ factor \citep[see][for details]{HallYork01}. {The first one is based on a polynomial approximation where after imposing a significance level $\alpha$, the correction factor $\lambda_{\alpha}$ is approximated with the following expression:}
\begin{equation}\label{lambdaalpha}
\lambda_{\alpha}=\frac{0.94029\alpha^3-1.59914\alpha^2 + 0.17695\alpha+ 0.48971}{\alpha^3-1.77793\alpha^2 + 0.36162\alpha + 0.42423}.
\end{equation}
{The second one uses Monte Carlo techniques considering a simple unimodal distribution. In particular, \cite{HallYork01} suggest to generate the resamples (of same sample size as the original data) obtained from a unimodal distribution resembling the sampled one and they claim that, in practice, normal distribution produce a good level accuracy. 

\cite{HallYork01} method should not be used in the general case of testing $k$--modality as the bootstrap test cannot be directly calibrated under this hypothesis, since it depends on the unknown quantities $f^{1/5}(t_i)/|f''(t_i)|^{2/5}$, where $t_i$ are the ordered turning points of $f$, with $i=1,\ldots,(2k-1)$. }

{As showed in \cite{Ameijeirasml}, the critical bandwidth of \cite{HallYork01} or \cite{Silverman81}, when $f$ has a bounded support, also plays a relevant role when the goal is to estimate the modes locations. When the true number of modes is known, under some general assumptions, the kernel density estimation with the critical bandwidth provides a good estimation of the modes and antimodes locations.}

A distribution estimation using the critical bandwidth of \cite{Silverman81} is also employed by \cite{FisMar01}, who considered the following Cram\'er--von Mises test statistic for testing $k$--modality,

\begin{equation}
T_k=\overset {n} {\underset {i=1} \sum} \left( \hat{F}_{h_k}(X_{(i)}) - \frac{2i-1}{2n} \right)^2 + \frac{1}{12n},
\label{fmstatistic}
\end{equation}
being $\hat{F}_{h_k}(x)=\int_{-\infty}^x {\hat{f}_{h_k}(t)dt}$. $H_0$ is rejected for large values of $T_k$ (\ref{fmstatistic}), whose distribution is approximated by a bootstrap algorithm, where resamples are generated from $\hat{f}_{h_k}$.

\subsection{Using an excess--mass statistic}
\label{background:excess}
\begin{figure}
\centering
\includegraphics[width=0.45\textwidth]{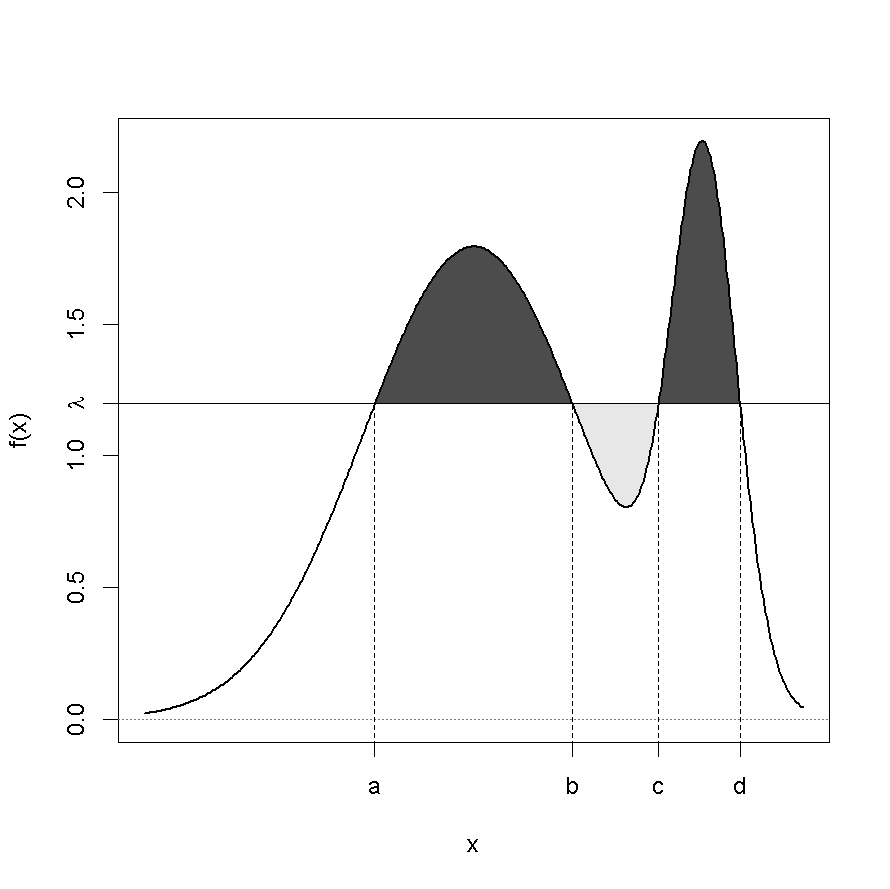}
 \caption{The excess mass for $k$ modes is the largest probability mass, exceeding a given level $\lambda$, when taking $k$ intervals. In this case, the excess mass for two modes is equal to the dark grey area (and obtained with the union of the intervals $[a,b]$ and $[c,d]$) and for one mode is equal to the dark grey minus the light grey area (and obtained with $[a,d]$). Then, in terms of excess mass, for the represented value of $\lambda$, the difference between assuming bimodality and unimodality is the light grey area.}
 \label{figexcmass}
\end{figure}

The identification of a mode in a density estimate by finding a \emph{significant} excess mass is the basic idea in the proposals by \cite{MulSaw91}, \cite{ChengHall98} and \cite{Ameijeirasml}. The empirical excess mass for $k$ modes and a constant $\lambda$ is defined as:

\begin{equation}
E_{n,k}(\mathbb{P}_n,\lambda)=\underset{C_1(\lambda),...,C_k(\lambda)}{\mbox{sup}} \left\{ \overset {k} {\underset {m=1} \sum} \mathbb{P}_n(C_m(\lambda))-\lambda ||C_m(\lambda)||\right\},
\label{elems}
\end{equation}  
where the supremum is taken over all families $\{C_m(\lambda) : m = 1, \cdots, k\}$ of closed intervals with endpoints at data points. $||C_m(\lambda)||$ denotes the measure of $C_m(\lambda)$ and $\mathbb{P}_n(C_m(\lambda))=(1/n) \sum_{i=1}^{n} \mathcal{I}(X_i \in C_m(\lambda))$, where $\mathcal{I}$ is the indicator function. The difference $D_{n,k+1}(\lambda)=E_{n,k+1}(\mathbb{P}_n,\lambda) - E_{n,k}(\mathbb{P}_n,\lambda)$ measures the plausibility of the null hypothesis, that is, large values of $D_{n,k+1}(\lambda)$ would indicate that $H_0$ is false. An example of the theoretical excess mass difference for a bimodal density is shown in Figure~\ref{figexcmass} for illustrative purposes. Using these differences, \cite{MulSaw91} defined as the excess mass statistic for testing $H_0:\;j=k$,

\begin{equation} 
\label{msstatistic}
\Delta_{n,k+1}=  \underset{\lambda}{\max } \{D_{n,k+1}(\lambda) \},
\end{equation}
Their proposal for testing unimodality is to calibrate this test statistic using a Monte Carlo calibration, where resamples are generated from the uniform distribution. The same approach was already proposed by \cite{Hartigan85} with the \textit{dip} unimodality test, since both quantities (dip and excess mass) coincide up to a factor.

The performance in practice of the calibration algorithm proposed for (\ref{msstatistic}) was remarkably conservative and \cite{ChengHall98} designed a bootstrap procedure for approximating the distribution of $\Delta_{n,2}$ under the hypothesis of unimodality generating the resamples from a family of parametric functions, guaranteeing an asymptotic correct behaviour. When analysing \cite{ChengHall98} proposal in simulated scenarios \citep[see][]{Ameijeirasml}, the calibration of the test was not satisfactory in the ``complicated'' unimodal models due the lack of flexibility of this parametric approach. Also, extending this test to the general case of testing $k$--modality is not an easy task. For those reasons, a completely nonparametric alternative for testing $H_0:\;j=k$ vs. $H_a:\;j>k$ has been proposed by \cite{Ameijeirasml}. Their method consist in calibrating the excess mass statistic given in (\ref{msstatistic}) using a bootstrap procedure, where the resamples are generated from (a modified version of) $\hat{f}_{h_k}$. The modification of the kernel density estimator ensures the correct calibration of this test, under some regularity conditions \citep[similar to those ones needed in][]{ChengHall98}. Although, in general, the \cite{Ameijeirasml} proposal presents a correct behaviour even when the sample size is ``small'' ($n=50$), when knowing the compact support $I$ where the modes and antimodes lie, the \cite{HallYork01} bandwidth can be employed (for generating the resamples), improving the results of this test.

When deciding which proposal should be chosen, it must be considered that an asymptotic correct behaviour is just expected in the unimodality tests of \cite{HallYork01} (when $f$ has a bounded support or when employing the compact support $I$) and \cite{ChengHall98} and in the multimodality test of \cite{Ameijeirasml}. A complete simulation study comparing all the aforementioned proposals is provided in \cite{Ameijeirasml}, showing that the other proposals \citep{Silverman81,FisMar01} for testing $H_0:\;j=k$ vs. $H_a:\;j>k$, when $k>1$, exhibit an unsatisfactory behaviour.

\section[Using multimode]{Using \pkg{multimode}}

A complete description of the \pkg{multimode} package capabilities is provided in this section. Specifically, the package includes the datasets and the functions shown in Table~\ref{functions}. First, the different datasets available in the package will be described. Second, the usage of different functions for exploring the number of modes will be illustrated. Finally, the functions for testing multimodality and estimating the location of modes and antimodes will be introduced.

\begin{table}
\begin{center}
\begin{tabular}{ll}
\hline
Dataset & Description\\
\hline
\code{acidity} & Acid--neutralizing capacity \\
\code{chondrite} & Percentage of silica in chondrite meteors \\
\code{enzyme} & Blood enzymatic activity \\
\code{galaxy} & Velocities of galaxies \\
\code{geyser} & Waiting time between geyser eruptions \\
\code{stamps} & Stamps thickness \\
\hline
Function & Description\\
\hline
\code{critbw} & Critical bandwidth computation\\
\code{excessmass} & Excess mass statistic\\
\code{locmodes} & Location of modes and antimodes\\
\code{modeforest} & Mode forest\\
\code{modetest}& Test for the number of modes\\
\code{modetree} & Mode tree\\
\code{nmodes} & Number of modes\\
\code{sizer} & SiZer map\\
\hline
\end{tabular}
\caption{Summary of \pkg{multimode} package contents.}
\label{functions}
\end{center}
\end{table}

\subsection{Data description}\label{data_description}

The package \pkg{multimode} includes some classical datasets for which determining the number of different groups in the sample and/or exploring the location of modes and antimodes are relevant issues. The first dataset, \code{acidity}, analysed by \cite{Crawford94}, contains, on the log scale, the Acid--Neutralizing Capacity (ANC) measured in a sample of 155 lakes in North--Central Wisconsin (USA). ANC describes the capability of a lake to absorb acid, where low ANC values may lead to a loss of biological resources. The dataset \code{chondrite}, included in Table~2 of \cite{GoodGask80}, gathers the percentage silica (in \%) in 22 chondrite meteors. The dataset \code{enzyme}, introduced by \cite{Bechtel93}, collects a sample with the distribution of enzymatic activity in the blood, for an enzyme involved in the metabolism of carcinogenic substances. The dataset \code{galaxy} provides the velocities in km/sec of different galaxies (diverging away from our own galaxy) from the unfilled survey of the Corona Borealis region. In this dataset introduced by \cite{Postman86} and further studied by \cite{Roeder90}, multimodality is an evidence for voids and superclusters in the universe. The dataset in \code{geyser} presents the interval times between the starts of the geyser eruptions observed during different periods in the Old Faithful Geyser in Yellowstone National Park, Wyoming, USA. The included periods are: October 1980, obtained from Table~3 of \cite{Hardle12} and the supplementary material of \cite{Weisberg05}; and August 1985, from Table~1 in \cite{Azzalini90}. Finally, the dataset \code{stamps}, analysed in \cite{IzenSom88}, consists of thickness measurements (in millimetres) of 485 unwatermarked used white wove stamps of the 1872 Hidalgo stamp issue of Mexico. All of them had an overprint with the year (1872 or either an 1873 or 1874) and some of them were watermarked (Papel Sellado or LA+-F), being this information also included inside \code{stamps}. Since the stamps value depends on its scarcity, it is of importance to determine the number of available groups in a particular stamp issue. For this particular stamp issue, although the watermark is some stamps (in 29 of 485) helps to conclude that there are at least two groups, the question about the number of groups can be answered analysing the underlying number of modes.

Some of these datasets (\code{acidity}, \code{enzyme}, \code{galaxy} or \code{stamps}) were used in the statistical literature for illustrating mixtures of parametric models. The nonparametric (both exploratory and inferential) tools included in \pkg{multimode} could be seen as a preliminary tool for determining the number of modes. Some references can be found in \cite{McLachlan00} or \cite{Richardson97}. In other datasets (\code{chondrite}, \code{geyser} or \code{stamps}), testing or exploring the number of modes is an important problem \textit{per se}. Some examples of their application can be found in \cite{ChMar99}, \cite{MulSaw91} or \citet[][Sect. 9.2]{Scott15}. In the subsequent sections, the \code{stamps} dataset will be used for illustrating the functions available in the \pkg{multimode} package.

\subsection[Exploring data]{Exploring data with \pkg{multimode}}\label{exploring_data}

When the objective is to explore the number of modes in a sample, a simple solution might be to observe the number of peaks in the kernel density estimation for different values of $h$. In order to facilitate this task, using the Gaussian kernel and a given bandwidth parameter \code{bw}, the function \code{nmodes} computes the estimated number of modes in the real line or in a support bounded by \code{lowsup} and \code{uppsup}. This kernel density estimation is calculated in \code{n} equally spaced points of the variable for computational reasons (as in the \code{density} function from the \pkg{stats} package). For instance, using the code below, it can be seen that the estimated number of modes using the rule of thumb and the plug--in rule (\code{bw.nrd0} and \code{bw.SJ} from the \pkg{stats} package and illustrated in Figure~\ref{figapmm2}, panel d) is, respectively, two and nine.

\begin{verbatim}
R> data(stamps)
R> bwRT <- bw.nrd0(stamps) ; bwPI <- bw.SJ(stamps)
R> nmodes(data=stamps,bw=bwRT,lowsup=-Inf,uppsup=Inf,n=2^15)
R> nmodes(data=stamps,bw=bwPI,lowsup=-Inf,uppsup=Inf,n=2^15)
\end{verbatim}

Based on the idea of exploring the number of modes (and their location) for different values of $h$, the three different graphical tools, presented in Section~\ref{exploratory_tools}, have been implemented in \pkg{multimode}: \code{modetree}, \code{modeforest} and \code{sizer}. The outputs from these exploratory functions and the arguments used for their computation are detailed below. The common characteristics, in the three of them, are: the exploratory features will be calculated in a finite number of grid points (the common argument is the first element of \code{gridsize}); the number of modes will be determined according to a value of $h$ and the employed bandwidth values can be chosen by the practitioner (\code{bws}, \code{cbw1}, \code{cbw2} and the second element \code{gridsize}); a graphical display is generated (or added to the current graphic) with different plot arguments (\code{display}, \code{logbw}, \code{xlab}, \code{ylab}); an output related with the modes locations is returned.  

The different exploratory tools (\code{modetree}, \code{modeforest} and \code{sizer}) include three options for providing the bandwidths. The first one is to use a range of bandwidth parameters in the argument \code{bws} and the exploratory tool is computed in a grid of $h$ between the given values and size equal to the second element of the argument \code{gridsize}. By default, a grid of size 151 is computed between a lower bandwidth equal to twice the distance between the grid points used for estimating the density and upper bandwidth equal to the data range. The second option considers the critical bandwidths for \code{cbw1} and \code{cbw2} modes as the range of bandwidths. The third method allows to include a vector of bandwidths in the argument \code{bws} with size greater than two. Then, these exploratory tools are represented (using a $\log_{10}$ scale for the bandwidths if \code{logbw=TRUE}) when the argument \code{display} is \code{TRUE} with the titles in the x and y axis provided by \code{xlab} and \code{ylab}, as usual. 

The mode tree introduced by \cite{MinSco93} and implemented in the function \code{modetree}, shows with continuous lines the estimated mode locations for each bandwidth. For \code{modetree}, the first element of \code{gridsize} is equal to the number of equally spaced points at which the density is to be estimated. Moreover, the mode tree can be added to another plot when the argument \code{addplot} is \code{TRUE}. Also, the color lines in the mode tree can be chosen with the argument \code{col.lines}. Below, an example with the code lines for computing the mode tree for the \code{stamps} dataset between the bandwidths $8\cdot10^{-4}$ and $8\cdot10^{-3}$ is shown (its representation appears in Figure~\ref{figapmm2}, panel b).

\begin{verbatim}
R> mtstamps <- modetree(data=stamps,bws=c(0.0008,0.008),
+    gridsize=c(512,151),cbw1=NULL,cbw2=NULL,display=TRUE,logbw=FALSE,
+    addplot=FALSE,xlab="Thickness (in mm)",ylab=NULL,col.lines="black")  
R> names(mtstamps)

[1] "locations"  "bandwidths"
\end{verbatim}

This function returns a list containing the following components: \code{locations}, a matrix with the estimated modes locations for each bandwidth; and \code{bandwidths}, the bandwidths employed for computing the mode tree. The plot and the argument \code{locations} returned by the function \code{modetree} can be useful for exploring where the different modes are located when the number of modes is not clear and a further insight on the data distribution is required. In this case, the principal mode appears between the values 0.0765 and 0.0793, the secondary mode between 0.0986 and 0.1011, and so forth.

The mode forest, introduced by \cite{Minetal98}, is provided by \code{modeforest}. This graphical tool is generated by looking simultaneously at a collection of mode trees generated by the original sample and \code{B} random resamples drawn with replacement from the original one. 

For the \code{modeforest} and \code{sizer}, the first element of \code{gridsize} is equal to the number of grid points in the horizontal (values of the variable) axis. In both cases, the horizontal values plotted are bounded by the interval (\code{from},\code{to}), being this interval equal to the data range by default. In the \code{modeforest} function, the number of equally spaced points at which the density is to be estimated is chosen by the argument \code{n}. The mode forest for the \code{stamps} dataset between the bandwidths $8\cdot10^{-4}$ and $8\cdot10^{-3}$ (represented in Figure~\ref{figapmm2}, panel c) can be obtained as follows:

\begin{verbatim}
R> mfstamps <- modeforest(data=stamps,bws=c(0.0008,0.008),
+    gridsize=c(100,151),B=99,n=512,cbw1=NULL,cbw2=NULL,display=TRUE,
+    logbw=FALSE,from=NULL,to=NULL,xlab="Thickness (in mm)",ylab=NULL)
R> names(mfstamps)

[1] "modeforest" "range.x"    "range.bws" 
\end{verbatim}

The output is a matrix \code{modeforest} including the percentage of times that an estimated mode falls in each location--bandwidth pixel. The functions \code{modeforest} and \code{sizer} return \code{range.x} (the employed location values to represent the mode forest or the SiZer map) and \code{range.bws} (the bandwidths used for computing the exploratory tool). In the \code{modeforest} plot, modes can be detected observing the dark grey vertical traces, but one should be careful with the very dark areas (as the one next to 0.06) since, due to the resampling algorithm, it is possible that spurious modes (created by some atypical data points) may seem visually more prominent than real modes \citep[as pointed out by][]{Minetal98}. Observing Figure~\ref{figapmm2} (panel c), the mode forest suggests at least seven modes for the \code{stamps} dataset.

With the \code{sizer} function the assessment of SIgnificant ZERo crossing of the derivative of the smoothed curve is computed for a given sample. In each location--bandwidth pixel, the SiZer map shows the significant features of the smoothed curve using, by default, the colours described in Section~\ref{exploratory_tools}, but they can be replaced using the \code{col.sizer} argument. For analysing the behaviour of the curve, the four quantile approximations proposed by \cite{ChMar99} are implemented in the \code{sizer} function using the argument \code{method}. The available quantiles are: the pointwise Gaussian quantiles ($q_1$), when \code{method=1}; approximate simultaneous over location $x$ Gaussian quantiles ($q_2$), when \code{method=2}; bootstrap quantile simultaneous over location $x$ ($q_3$), when \code{method=3}; and bootstrap quantile simultaneous over (location and bandwidth) $x$ and $h$ ($q_4$), when \code{method=4}. Bootstrap quantiles $q_3$ and $q_4$ are computed generating \code{B} random samples drawn with replacement from the sample. In methods $q_2$, $q_3$ and $q_4$; grey colour (by default) is employed when the Effective Sample Size in (\ref{ESSsizer}) is less than the value \code{n0}. A legend indicating the meaning of the different colours is also provided in the plot position given in \code{poslegend} when the argument \code{addlegend} is \code{TRUE}. The different SiZer maps for the \code{stamps} dataset between the bandwidths $8\cdot10^{-4}$ and $0.02$ (represented in Figure~\ref{figapmm2}; panels e, f, h and i) can be obtained as shown below (varying the value of \code{method} between 1 and 4). For computing the quantiles $q_2$, $q_3$ and $q_4$ it was taken \code{n0=5} and the number of bootstrap replicas in methods $q_3$ and $q_4$ is \code{B=500}.

\begin{verbatim}
R> sizerstamps <- sizer(data=stamps,method=1,bws=c(0.0008,0.02),
+    gridsize=NULL,alpha=0.05,B=NULL,n0=NULL,cbw1=NULL,cbw2=NULL,
+    display=TRUE,logbw=TRUE,from=NULL,to=NULL,col.sizer=NULL,
+    xlab="Thickness (in mm)",ylab=NULL,addlegend=TRUE,
+    poslegend="topright")
R> names(sizerstamps)

[1] "sizer"     "lower.CI"  "estimate"  "upper.CI"  "ESS"      
[6] "range.x"   "range.bws"
\end{verbatim}

Apart from the already described arguments, \code{sizer} returns a list with five matrices containing different information in each location--bandwidth pixel: \code{sizer}, {with the significant behaviours of the smoothed curve in each location--bandwidth pixel (1: significantly decreasing, 2: not significantly different from zero, 3: significantly increasing or 4: low data for meaningful inference)}; \code{lower.CI} with the lower limits of the confidence interval, $\mbox{CI}^{-}(x,h)$; \code{estimate}, with the derivative values of the kernel density estimation, $\hat{f}_{h}'(x)$; \code{upper.CI} with the upper limits of the confidence interval, $\mbox{CI}^{+}(x,h)$; and \code{ESS}, with the Effective Sample Size. 

As noted before, in the SiZer maps (represented in Figure~\ref{figapmm2}; panels e, f, h and i), by default, blue colour indicates locations where, for a given bandwidth, the smoothed curve is significantly increasing, red colour shows where it is significantly decreasing and orchid indicates where it is not significantly different from zero. Then, focusing on $\log_{10}(h)$ values, modes can be detected by blue--red patterns. In this case, the SiZer maps computed with Gaussian quantiles just detect, at most, one mode around 0.08. The conclusion with those ones constructed with bootstrap confidence intervals vary with the bandwidth. For all the bandwidth values, both methods capture a principal mode before the value 0.08 and for several bandwidth parameters is also detected a secondary mode around 0.10. The third and the fourth mode (around 0.09 and 0.11) that appears in the mode tree (Figure~\ref{figapmm2}, panel b) are only significant modes for some bandwidth parameters for $q_3$. Finally, both methods, $q_3$ and $q_4$, detect another mode near 0.07 for some bandwidth values. Then, depending on the bandwidth parameter, the conclusion using the quantile $q_3$ is that there are between one and five modes (in order of appearance, around 0.08, 0.10, 0.09, 0.11 and 0.07), while $q_4$ detects between one and three modes (around 0.08, 0.10 and 0.07).

\subsection[Testing and locating modes]{Testing and locating modes with \pkg{multimode}}

The \pkg{multimode} package has implemented all the test presented in the Section~\ref{testing_procedures}. In particular, it allows to compute the critical bandwidth of \cite{Silverman81} and \cite{HallYork01} (with the function \code{bw.crit}) and the excess mass of \cite{MulSaw91} (with \code{excessmass}). Their associated p--values can be also obtained, with \code{modetest}, using different testing proposals. For the three functions (\code{bw.crit}, \code{excessmass} and \code{modetest}), the investigated number of modes can be specified in the argument \code{mod0}. 

For \code{bw.crit} and for the testing proposals using the critical bandwidth in \code{modetest}, when the compact support is unknown, the critical bandwidth introduced by \cite{Silverman81} is computed and if the finite values of the support limits are provided (via arguments \code{lowsup} and \code{uppsup}) the one proposed by \cite{HallYork01} is calculated. Both arguments should be used in \code{modetest} when employing the \cite{HallYork01} proposal or for computing the new proposal when the compact support is known (see Section~\ref{background:excess}). As in the \code{nmodes} function, the number of equally spaced points at which the density is to be estimated is chosen by the argument \code{n}. Since a dichotomy method is employed for computing the critical bandwidth, the parameter \code{tol} is used to determine a stopping time in such a way that the error committed in the computation of the critical bandwidth is less than \code{tol}.

For \code{excessmass} and in the testing proposals using the excess mass in \code{modetest}, when there are repeated data in the sample or the distance between different pairs of data points shows ties, a data perturbation is applied. This modification is made in order to avoid the induced discretization of the data which has important effects on the computation of this test statistic. The perturbed sample is obtained by adding a sample from the uniform distribution in the support minus/plus a half of the minimum of the positive distances between two sample points.

Since the excess mass for one mode is twice the dip, this equality can be used for a ``fast'' computation of the excess mass for one mode. When \code{mod0} is greater than one and the sample size is ``large'', a two--steps approximation (\code{approximate=TRUE}) can be performed in order to improve the computational time efficiency. This two--steps approximation is achieved creating two grids of values of size the elements in \code{gridsize}. First, since the possible $\lambda$ candidates for maximizing $D_{n,k+1}(\lambda)$ can be directly obtained from the $C_m(\lambda)$ sets that could maximize $E_{n,k+1}$ and $E_{n,k}$ \citep[see Supplementary Material in][]{Ameijeirasml}, the possible values of $\lambda$ are computed by looking to the empirical excess mass function in some endpoints candidates for $C_m(\lambda)$ (the number of employed points is equal to the first element of \code{gridsize}) and also in the $\lambda$ values associated to the empirical excess mass for one mode. Once a $\lambda$ maximizing the approximated values of $D_{n,k+1}(\lambda)$ is chosen, in order to obtain the approximation of the excess mass test statistic, in its neighbourhood, a grid of possible $\lambda$--values is created, being its length equal to the second element of \code{gridsize}, and the exact value of $D_{n,k+1}(\lambda)$ is calculated for these values of $\lambda$ \citep[using the algorithm proposed by][]{MulSaw91}.

An illustration with the \code{stamps} dataset is shown below. First, the critical bandwidth of \cite{Silverman81} and \cite{HallYork01}, in the interval $I=[0.04,0.15]$, is computed for two modes. Second, the exact and approximated version of the excess mass test statistic of \cite{MulSaw91} for two modes are obtained.

\begin{verbatim}
R> bw.crit(data=stamps,mod0=2,lowsup=-Inf,uppsup=Inf,n=2^15,tol=10^(-5))
R> bw.crit(data=stamps,mod0=2,lowsup=0.04,uppsup=0.15,n=2^15,tol=10^(-5))
R> excessmass(data=stamps,mod0=2,approximate=FALSE)
R> excessmass(data=stamps,mod0=2,approximate=TRUE,gridsize=c(20,20))
\end{verbatim}

Once the different test statistics are computed, the number of modes for the underlying density of a given sample can be tested with the function \code{modetest}. The different proposals that can be used for testing the number of modes (using the argument \code{method}) are those ones introduced in Section~\ref{testing_procedures}. The available methods, based on the critical bandwidth (see Section~\ref{background:critical}), include: \cite{Silverman81} (\code{SI}), \cite{HallYork01} (\code{HY}) and \cite{FisMar01} (\code{FM}). Based on the excess mass (Section~\ref{background:excess}): \cite{Hartigan85} \citep[\code{HH}, equivalent to the proposal of][]{MulSaw91}, \cite{ChengHall98} (\code{CH}) and the new proposal of \cite{Ameijeirasml} (\code{ACR}) is also included. For calculating the corresponding p--value, all the available proposals require bootstrap or Monte Carlo resamples and the number of replicates can be specified with the argument \code{B}.

For \code{SI}, \code{HY} and \code{ACR} proposals, the argument \code{submethod} is available. In the \code{SI} case, two resampling methods are implemented: when \code{submethod=1}, the resamples are generated from the rescaled bootstrap resamples as proposed by \cite{Silverman81} (see Section~\ref{background:critical}); if \code{submethod=2}, the resamples are generated from $\hat{f}_{h_k}$. In the \code{ACR} method, the approximated version of the excess mass can be employed, for computational time efficiency reasons, by setting \code{submethod=2}; if \code{submethod=1}, then the exact value of the excess mass test statistic is computed.

As pointed out in Section~\ref{testing_procedures}, the bounded support (\code{lowsup} and \code{uppsup}) is necessary when the \cite{HallYork01} proposal (\code{HY}) is employed and $f$ has not a compact support and it can be also used with the \code{ACR} proposal. In the \code{ACR} case, the parameter \code{tol2} is the accuracy required in the integration of the calibration function when the compact support is known \citep[see][]{Ameijeirasml}. {As mentioned in Section~\ref{background:critical}, a level correction (achieved with the $\lambda_{\alpha}$ factor) is needed in the bootstrap procedure of \cite{HallYork01}. The two suggested approximations for its computation are provided in the \code{HY} test using the argument \code{submethod}. The \code{submethod} 1 corresponds with the asymptotic correction of \cite{Silverman81} test based on the limiting distribution of the test statistic, i.e. it consists in using equality (\ref{lambdaalpha}). In equation (\ref{lambdaalpha}), since the value of $\lambda_{\alpha}$ depends on $\alpha$, when \code{submethod=1}, the significance level must be previously determined with \code{alpha}. The \code{submethod} 2 is based on Monte Carlo techniques where the resamples are generated from the normal distribution. For this reason, when \code{submethod=2}, the number of normal--distributed samples (\code{nMC}) and the number of bootstrap resamples (\code{BMC}) used for computing the p--value in each Monte Carlo sample are needed.} 

Finally, the \code{modetest} function includes the argument \code{full.result}. When this argument equals \code{TRUE}, the function returns a list with both, the test statistic (\code{statistic}) and the associated p--value (\code{p.value}); when it is \code{FALSE}, just the \code{p.value} is returned.

The different p--values obtained for the \code{stamps} dataset with the \cite{Ameijeirasml} proposal (calculating the exact value of the excess mass) are reproduced in Table~\ref{tabapmm2} and they can be obtained as follows (varying the value of \code{mod0} between 1 and 9):

\begin{verbatim}
R> modeteststamps <- modetest(data=stamps,mod0=1,method="ACR",B=500,
+    full.result=TRUE,submethod=1,n=2^10,tol=10^(-5))
R> names(modeteststamps)

[1] "p.value"   "statistic"
\end{verbatim}

Assuming that the compact support for the \code{stamps} dataset is $I=[0.04,0.15]$ \citep[see][]{IzenSom88}, the modification of the \cite{Ameijeirasml} proposal with known compact support can be obtained as follows

\begin{verbatim}
R> modetest(data=stamps,mod0=1,method="ACR",B=500,full.result=FALSE,
+    submethod=1,lowsup=0.04,uppsup=0.15,n=2^10,tol=10^(-5),tol2=10^(-5))
\end{verbatim}

The p--values of the other proposals allowing for testing a general number of modes (\code{SI} and \code{FM}) are obtained with the below code lines (varying the value of \code{mod0} between 1 and 9). 

\begin{verbatim}
R> modetest(data=stamps,mod0=1,method="SI",B=500,full.result=FALSE,
+    submethod=1,n=2^10,tol=10^(-5))
R> modetest(data=stamps,mod0=1,method="FM",B=500,full.result=FALSE,
+    n=2^10,tol=10^(-5))
\end{verbatim}

The other critical bandwidth based method, \code{HY}, should only be used for testing unimodality when when $f$ has a bounded support or when the modes and antimodes lie in a known closed interval $I$, in this case $I=[0.04,0.15]$. The test with both alternatives for approximating the $\lambda_\alpha$: a first approach based on a polynomial approximation (\code{submethod=1}) and a second option using Monte Carlo techniques (\code{submethod=2}), can be computed as follows:  

\begin{verbatim}
R> modetest(data=stamps,method="HY",B=500,full.result=FALSE,lowsup=0.04,
+    uppsup=0.15,n=2^10,tol=10^(-5),submethod=1,alpha=0.05)
R> modetest(data=stamps,method="HY",B=500,full.result=FALSE,lowsup=0.04,
+    uppsup=0.15,n=2^10,tol=10^(-5),submethod=2,nMC=100,BMC=100)
\end{verbatim}

The p--values of the unimodality test based on the excess mass (\code{HH} and \code{CH}) can be obtained with the following code lines:

\begin{verbatim}
R> modetest(data=stamps,method="HH",B=500,full.result=FALSE)
R> modetest(data=stamps,method="CH",B=500,full.result=FALSE)
\end{verbatim}

Table~\ref{tabapmm1} shows the \code{p.values} obtained for all the unimodality tests available. {Note that, in the \code{ACR} case, \code{submethod=2} was not employed as when \code{mod0=1} the exact version of the excess mass is computed in a more efficient way, then it does not make sense to use its approximated version}. For all of them the null hypothesis of unimodality is rejected for a significance level $\alpha=0.05$. 

\begin{table}
\centering
\begin{tabular}{|c |c  | c  | c |c|c |c  | c |c|}
  \hline
\code{method} & \multicolumn{2}{c|}{\code{SI}} & \multicolumn{2}{c|}{\code{HY}} & \code{FM} & \code{HH} & \code{CH} & \code{ACR}  \\ \hline
\code{submethod} & 1 & 2 & 1 & 2 &  &  &  & 1 \\   \hline
P--value & 0.018 & 0.006 & 0 & 0 & 0 & 0.030 & 0 & 0 \\ 
   \hline
\end{tabular}
\caption{P--value obtained using different proposals for testing unimodality, with $B=500$ resamples.{ The employed testing procedures are: \code{SI} (using the rescaled, \code{submethod} 1, and the non--rescaled, \code{submethod} 2, bootstrap resamples), \code{HY} (using the two suggested approximations of $\lambda_{\alpha}$), \code{FM}, \code{HH}, \code{CH} and \code{ACR} (employing the exact version of the excess mass test statistic).}}
\label{tabapmm1}
\end{table}
The results for the tests (\code{SI}, \code{FM} and \code{ACR}) that allow testing $k$--modality, with $k>1$, are displayed in Table~\ref{tabapmm2} (with $k$ between one and nine). In the case of the \code{FM} proposal, for reproducing the \cite{FisMar01} results, the \code{stamps} data were also perturbed as done with the \code{excessmass} function. Similar results are obtained for the \code{SI} proposal, with and without data perturbation when using \code{submethod=1} or \code{submethod=2}; and for the \code{ACR} proposal, independently of using or not the known support $I=[0.04,0.15]$. Fixing a significance level $\alpha=0.05$, there is not a clear conclusion when using \code{SI} and \code{FM}. In the \code{SI} case the null hypothesis is not rejected for $k=2,3,7,8,9$ and for the \code{FM}, using the perturbed sample, it is not rejected just for $k=7$. While, in the single proposal that is well calibrated \citep[\code{ACR}, see][]{Ameijeirasml}, the null hypothesis is rejected until $k=3$ and it is not for $k \geq 4$, suggesting that the number of modes is equal to 4.

\begin{table}
\begin{center}
\begin{tabular}{ |c  | c c c c c c c c c| }
\hline
 $k$ & 1 & 2 & 3 & 4 & 5 & 6 & 7 & 8 & 9 \\ \hline
\code{SI}  & 0.018 & 0.394 & 0.090 & 0.008 & 0.002 & 0.002 & 0.488 & 0.346 & 0.614 \\  \hline
\code{FM}  & 0 & 0.006 & 0 & 0 & 0 & 0 & 0 & 0 & 0 \\
\code{FM}* & 0 & 0 & 0 & 0 & 0 & 0 & 0.096 & 0.014 & 0.046 \\ \hline
\code{ACR}* & 0 & 0.022 & 0.004 & 0.506 & 0.574 & 0.566 & 0.376 & 0.886 & 0.808 \\ \hline
\end{tabular}
\caption{P--value obtained using different proposals for testing $k$--modality, with $k$ between 1 and 9, employing $B=500$ resamples. {The employed testing procedures are: \code{SI} over the original sample (using the rescaled bootstrap resamples), \code{FM} over the original sample, \code{FM} over the perturbed sample (\code{FM}*) and \code{ACR} over the perturbed sample (employing the exact version of the excess mass test statistic).}}
\label{tabapmm2}
\end{center}
\end{table}

Once the number of modes is known, the function \code{locmodes} provides the estimation of the locations of modes and antimodes and their estimated density value. In this case, the compact support of the variable (which is known) can be used to obtain a good estimator of the modes and antimodes locations (see Section~\ref{background:critical}). In other scenarios, one should be careful about the conclusions as the critical bandwidth of \cite{Silverman81} may create artificial modes in the tails \citep[see][]{HallYork01}. 

The arguments for \code{locmodes} function include those ones mentioned in the \code{bw.crit} function: \code{mod0}, \code{lowsup}, \code{uppsup}, \code{n} and \code{tol}. It also allows the representation of the estimation (for the number of modes indicated in \code{mod0}) of the density, modes and antimodes with the argument \code{display}. The remaining graphical arguments (\code{addplot}, \code{xlab}, \code{ylab}, \code{addLegend}, \code{posLegend}) were already described in the \code{modetree} and \code{sizer} functions. 

\begin{figure}
 \centering
\subfloat{
    \includegraphics[width=0.49\textwidth]{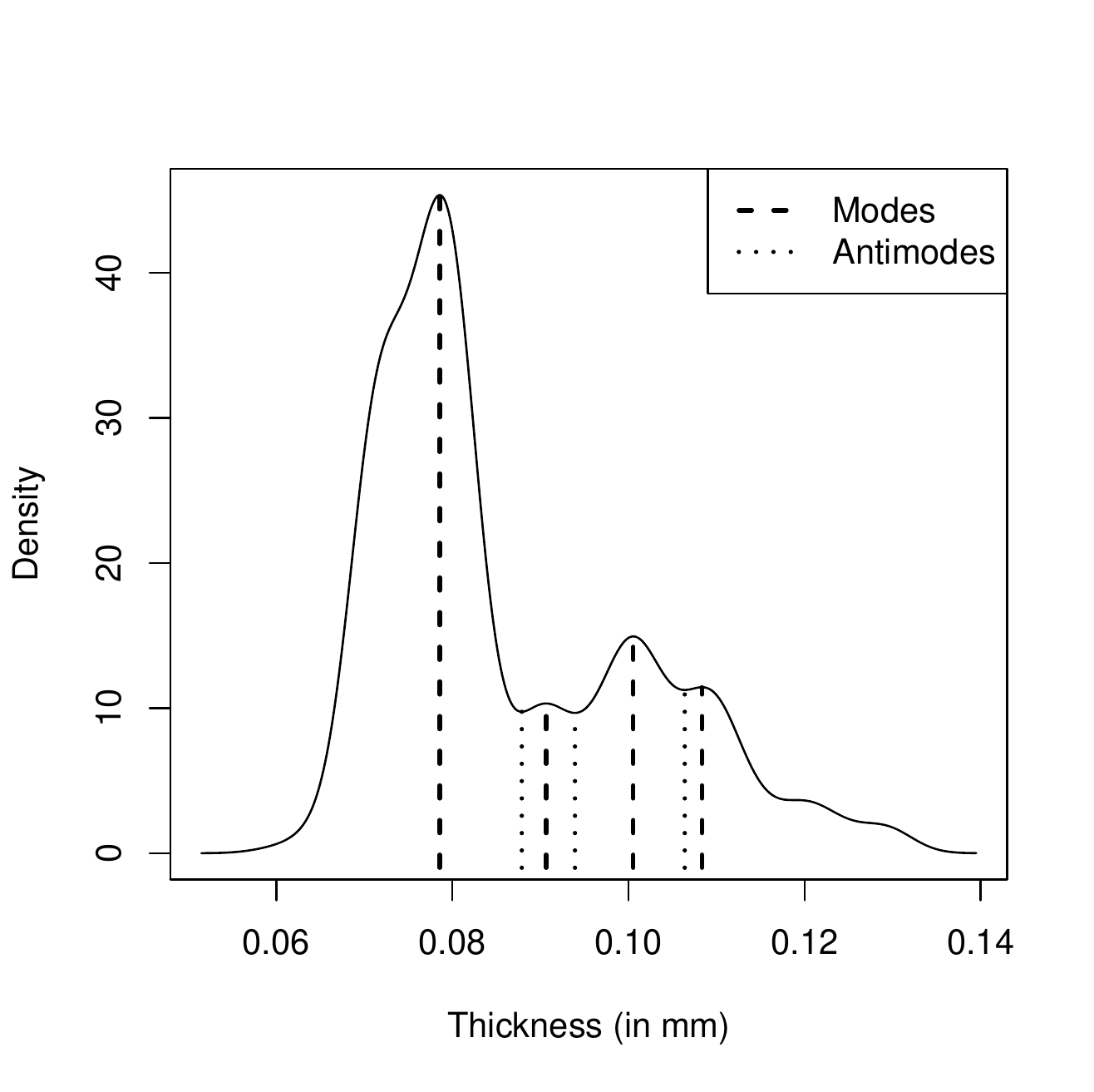}}
\subfloat{
    \includegraphics[width=0.49\textwidth]{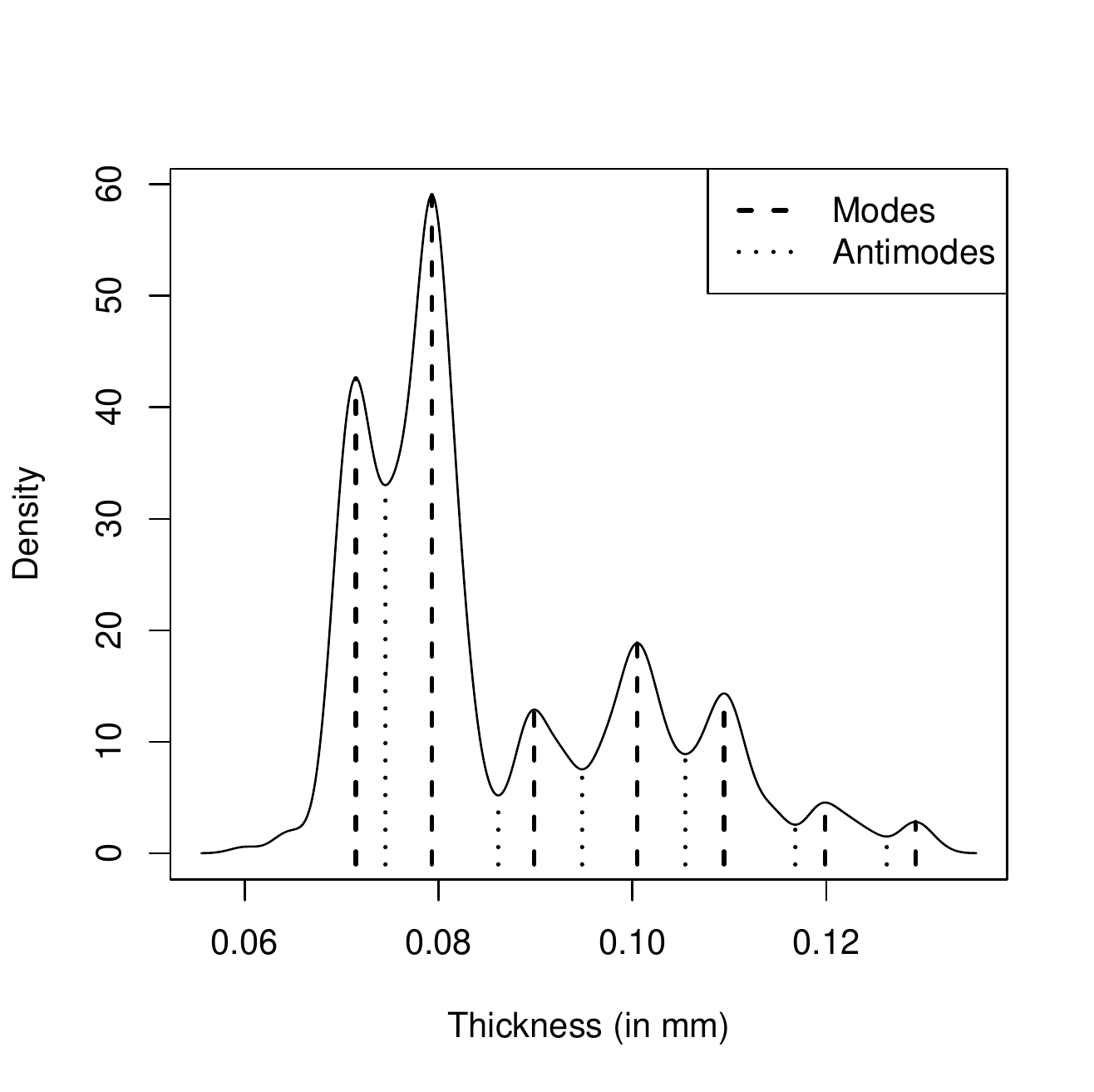}} \\
 \caption{Estimation of the density, modes and antimodes for the sample of 485 stamps from the 1872 Hidalgo Issue of Mexico, obtained with the function \code{locmodes} for \code{mod0=4} (left) and \code{mod0=7} (right) modes.}
 \label{figapmm3}
\end{figure}

The estimation of the modes and antimodes locations and their density value, assuming four \citep[\code{mod0=4},][]{Ameijeirasml} and seven \citep[\code{mod0=7},][]{IzenSom88} modes, can be obtained as follows (their representation is provided in Figure~\ref{figapmm3}):

\begin{verbatim}
R> lms <- locmodes(data=stamps,mod0=4,lowsup=0.04,uppsup=0.15,n=2^15,
+    tol=10^(-5),display=TRUE,addplot=FALSE,xlab="Thickness (in mm)",
+    ylab=NULL,addLegend=TRUE,posLegend="topright")
R> names(lms)

[1] "locations" "fvalue"    "cbw" 
\end{verbatim}

This function returns \code{locations}, a vector with the estimated locations of modes (odd positions of the vector) and antimodes (even positions); \code{fvalue}, a vector with their estimated density values; and \code{cbw}, the critical bandwidth of the sample for \code{mod0} modes. Regarding the obtained results assuming that the distribution has four modes, the estimated modes (odd positions of \code{locations}) are: 0.07857, 0.09065, 0.1006 and 0.1083. 

The results obtained after applying the \code{modetest} function can be helpful for having a better interpretation of the SiZer map (Figure~\ref{figapmm2}, panels e, f, h and i). If the conclusion is that there are four modes, the most plausible results are obtained with the bootstrap quantiles $q_3$ and, in that case, the estimated modes (in \code{locmodes}) coincide with those ones observed when a value of $\log_{10}(h)$ close to $-2.7$ is taken.

\section[Discussion]{Discussion}\label{discussion}

The available functions of the \proglang{R} package \pkg{multimode} were described in this paper. This package was developed with the objective of making the mode testing and exploring procedures, for linear data, accessible for the scientific community, and therefore, enabling its use in practical problems. As pointed out in Section~\ref{intro}, there are many examples in different disciplines where the identification of the number (and location) of modes is important \emph{per se}, or as a previous step for applying other procedures. Package \pkg{multimode} contains nonparametric graphical tools for (visually) exploring the number of modes and their estimated location and also testing proposals for determining the number of modes in the data distribution.

Up to the author's knowledge, \pkg{multimode} is the only statistical package that allows for testing, in a nonparametric way, a general number of modes and, also, it is the only one providing a well--calibrated method for testing unimodality. Obtaining a final p--value, instead of a graphical tool, can be useful when the objective is, e.g. to obtain conclusions about the number of modes in a systematic manner. This is the case of \cite{mcquillan14} or \cite{Joubert16} where they performed several times the unimodality test of \cite{Hartigan85}, dividing the sample, in the first case, in a temperature bin, and in the second case, in a collection of different CpGs (see Section~\ref{intro}). The combination of this package with other False Discovered Rate techniques (see, e.g. \code{p.adjust} from the \pkg{stats} package) allows to account for the multiple testing problem when the objective is to determine the number of modes.  

So far, \pkg{multimode} includes just exploratory and testing procedures for mode assessment for real random variables. However, the ideas in \cite{Ameijeirasml} can be extended to settings where there is a natural nonparametric estimator. This is the case with circular random variables, for instance. As mentioned before, in \proglang{R}, there are already some packages allowing for exploring the number of modes in this setting, such as the circular version of the SiZer map implemented in \pkg{NPCirc} \citep[see][]{Oliveira14b}. Referring to the testing approach, \cite{FisMar01} already introduced a proposal for determining the number of modes in this circular setting. In particular, they suggested to use the circular version of the $T_k$ test statistic, namely the $U^2$ of \cite{Watson61}. Future extensions of the \code{multimode} package could include some procedures for assessing the number of modes in other settings, such as the mentioned proposal of \cite{FisMar01}.

\section[Acknowledgements]{Acknowledgements}

The authors gratefully acknowledge the support of Project MTM2016--76969--P from the Spanish State Research Agency (AEI) and the European Regional Development Fund (ERDF), IAP network from Belgian Science Policy. Work of J. Ameijeiras-Alonso has been supported by the PhD Grant BES--2014--071006 from the Spanish Ministry of Economy, Industry and Competitiveness.



\bibliographystyle{apa-Jose}
\bibliography{references_multimode}

\newcommand{\noop}[1]{}
\begin{thebibliography}{45}
\expandafter\ifx\csname natexlab\endcsname\relax\def\natexlab#1{#1}\fi
\expandafter\ifx\csname url\endcsname\relax
  \def\url#1{{\tt #1}}\fi
\expandafter\ifx\csname urlprefix\endcsname\relax\def\urlprefix{URL }\fi

\bibitem[{Ameijeiras-Alonso \emph{et~al.}(2016)Ameijeiras-Alonso, Crujeiras, \&
  Rodr{\'\i}guez-Casal}]{Ameijeirasml}
Ameijeiras-Alonso, J., Crujeiras, R.~M., and Rodr{\'\i}guez-Casal, A. (2016).
\newblock Mode testing, critical bandwidth and excess mass.
\newblock {\em arXiv preprint arXiv:1609.05188\/}.
\newblock Submitted.

\bibitem[{Ameijeiras-Alonso \emph{et~al.}(2018)Ameijeiras-Alonso, Crujeiras, \&
  Rodr{\'i}guez-Casal}]{Ameijeiraspkg}
Ameijeiras-Alonso, J., Crujeiras, R.~M., and Rodr{\'i}guez-Casal, A. (2018).
\newblock {\em multimode: Mode Testing and Exploring\/}.
\newblock R package version 1.1.
\urlprefix\url{https://CRAN.R-project.org/package=multimode}

\bibitem[{Azzalini and Bowman(1990)Azzalini \& Bowman}]{Azzalini90}
Azzalini, A., and Bowman, A.~W. (1990).
\newblock A look at some data on the Old Faithful geyser.
\newblock {\em Applied Statistics\/}, {\bf 39\/}, 357--365.

\bibitem[{Bechtel \emph{et~al.}(1993)Bechtel, Bonaiti-Pellie, Poisson,
  Magnette, \& Bechtel}]{Bechtel93}
Bechtel, Y.~C., Bonaiti-Pellie, C., Poisson, N., Magnette, J., and Bechtel,
  P.~R. (1993).
\newblock A population and family study N-acetyltransferase using caffeine
  urinary metabolites.
\newblock {\em Clinical Pharmacology \& Therapeutics\/}, {\bf 54\/}, 134--141.

\bibitem[{Benaglia \emph{et~al.}(2009)Benaglia, Chauveau, Hunter, \&
  Young}]{Benaglia09}
Benaglia, T., Chauveau, D., Hunter, D.~R., and Young, D. (2009).
\newblock {mixtools}: An {R} Package for Analyzing Finite Mixture Models.
\newblock {\em Journal of Statistical Software\/}, {\bf 32\/}, 1--29.
\urlprefix\url{http://www.jstatsoft.org/v32/i06/}

\bibitem[{Chaudhuri and Marron(1999)Chaudhuri \& Marron}]{ChMar99}
Chaudhuri, P., and Marron, J.~S. (1999).
\newblock SiZer for Exploration of Structures in Curves.
\newblock {\em Journal of the American Statistical Association\/}, {\bf 94\/},
  807--823.

\bibitem[{Cheng and Hall(1998)Cheng \& Hall}]{ChengHall98}
Cheng, M.~Y., and Hall, P. (1998).
\newblock Calibrating the excess mass and dip tests of modality.
\newblock {\em Journal of the Royal Statistical Society. Series B\/}, {\bf
  60\/}, 579--589.

\bibitem[{Colombo \emph{et~al.}(2015)Colombo, Franzoni, \&
  Rossi-Lamastra}]{colombo15}
Colombo, M.~G., Franzoni, C., and Rossi-Lamastra, C. (2015).
\newblock Internal social capital and the attraction of early contributions in
  crowdfunding.
\newblock {\em Entrepreneurship Theory and Practice\/}, {\bf 39\/}, 75--100.

\bibitem[{Crawford(1994)}]{Crawford94}
Crawford, S.~L. (1994).
\newblock An application of the Laplace method to finite mixture distributions.
\newblock {\em Journal of the American Statistical Association\/}, {\bf 89\/},
  259--267.

\bibitem[{D{\"u}mbgen and Walther(2008)D{\"u}mbgen \& Walther}]{dumbgen2008}
D{\"u}mbgen, L., and Walther, G. (2008).
\newblock Multiscale inference about a density.
\newblock {\em The Annals of Statistics\/}, {\bf 36\/}, 1758--1785.

\bibitem[{Duong \emph{et~al.}(2008)Duong, Cowling, Koch, \& Wand}]{duong08}
Duong, T., Cowling, A., Koch, I., and Wand, M.~P. (2008).
\newblock Feature significance for multivariate kernel density estimation.
\newblock {\em Computational Statistics \& Data Analysis\/}, {\bf 52\/},
  4225--4242.

\bibitem[{Duong and Wand(2015)Duong \& Wand}]{Duong15}
Duong, T., and Wand, M. (2015).
\newblock {\em feature: Local Inferential Feature Significance for Multivariate
  Kernel Density Estimation\/}.
\newblock R package version 1.2.13.
\urlprefix\url{https://CRAN.R-project.org/package=feature}

\bibitem[{Fisher and Marron(2001)Fisher \& Marron}]{FisMar01}
Fisher, N.~I., and Marron, J.~S. (2001).
\newblock Mode testing via the excess mass estimate.
\newblock {\em Biometrika\/}, {\bf 88\/}, 419--517.

\bibitem[{Freeman and Dale(2013)Freeman \& Dale}]{freeman13}
Freeman, J.~B., and Dale, R. (2013).
\newblock Assessing bimodality to detect the presence of a dual cognitive
  process.
\newblock {\em Behavior research methods\/}, {\bf 45\/}, 83--97.

\bibitem[{Godtliebsen \emph{et~al.}(2002)Godtliebsen, Marron, \&
  Chaudhuri}]{godtliebsen02}
Godtliebsen, F., Marron, J., and Chaudhuri, P. (2002).
\newblock Significance in scale space for bivariate density estimation.
\newblock {\em Journal of Computational and Graphical Statistics\/}, {\bf
  11\/}, 1--21.

\bibitem[{Good and Gaskins(1980)Good \& Gaskins}]{GoodGask80}
Good, I.~J., and Gaskins, R.~A. (1980).
\newblock Density estimation and bump-hunting by the penalized likelihood
  method exemplified by scattering and meteorite data.
\newblock {\em Journal of the American Statistical Association\/}, {\bf 75\/},
  42--56.

\bibitem[{Hall and York(2001)Hall \& York}]{HallYork01}
Hall, P., and York, M. (2001).
\newblock On the calibration of Silverman's test for multimodality.
\newblock {\em Statistica Sinica\/}, {\bf 11\/}, 515--536.

\bibitem[{H{\"a}rdle(2012)}]{Hardle12}
H{\"a}rdle, W. (2012).
\newblock {\em Smoothing Techniques: with Implementation in S\/}.
\newblock Springer Science \& Business Media, New York.

\bibitem[{Hartigan and Hartigan(1985)Hartigan \& Hartigan}]{Hartigan85}
Hartigan, J.~A., and Hartigan, P.~M. (1985).
\newblock The Dip Test of Unimodality.
\newblock {\em Annals of Statistics\/}, {\bf 13\/}, 70--84.

\bibitem[{Izenman and Sommer(1988)Izenman \& Sommer}]{IzenSom88}
Izenman, A.~J., and Sommer, C.~J. (1988).
\newblock Philatelic mixtures and multimodal densities.
\newblock {\em Journal of the American Statistical Association\/}, {\bf 83\/},
  941--953.

\bibitem[{Joubert \emph{et~al.}(2016)Joubert, Felix, Yousefi, Bakulski, Just,
  Breton, Reese, Markunas, Richmond, Xu, K\"upers, Oh, Hoyo, Gruzieva,
  S\"oderh\"all, Salas, Ba\"iz, Zhang, Lepeule, Ruiz, Ligthart, Wang, Taylor,
  Duijts, Sharp, Jankipersadsing, Nilsen, Vaez, Fallin, Hu, Litonjua,
  Fuemmeler, Huen, Kere, Kull, Munthe-Kaas, Gehring, Bustamante,
  Saurel-Coubizolles, Quraishi, Ren, Tost, Gonzalez, Peters, H\r{a}berg, Xu,
  van Meurs, Gaunt, Kerkhof, Corpeleijn, Feinberg, Eng, Baccarelli, Neelon,
  Bradman, Merid, Bergstr\"om, Herceg, Hernandez-Vargas, Brunekreef, Pinart,
  Heude, Ewart, Yao, Lemonnier, Franco, Wu, Hofman, McArdle, {Van~der~Vlies},
  Falahi, Gillman, Barcellos, Kumar, Wickman, Guerra, Charles, Holloway,
  Auffray, Tiemeier, Smith, Postma, Hivert, Eskenazi, Vrijheid, Arshad, Ant\'o,
  Dehghan, Karmaus, Annesi-Maesano, Sunyer, Ghantous, Pershagen, Holland,
  Murphy, DeMeo, Burchard, Ladd-Acosta, Snieder, Nystad, Koppelman, Relton,
  Jaddoe, Wilcox, Mel\'en, \& London}]{Joubert16}
Joubert, B., Felix, J., Yousefi, P., Bakulski, K., Just, A., Breton, C., Reese,
  S.~E., Markunas, C., Richmond, R., Xu, C.-J., K\"upers, L., Oh, S., Hoyo, C.,
  Gruzieva, O., S\"oderh\"all, C., Salas, L., Ba\"iz, N., Zhang, H., Lepeule,
  J., Ruiz, C., Ligthart, S., Wang, T., Taylor, J., Duijts, L., Sharp, G.,
  Jankipersadsing, S., Nilsen, R., Vaez, A., Fallin, M., Hu, D., Litonjua, A.,
  Fuemmeler, B., Huen, K., Kere, J., Kull, I., Munthe-Kaas, M., Gehring, U.,
  Bustamante, M., Saurel-Coubizolles, M., Quraishi, B., Ren, J., Tost, J.,
  Gonzalez, J., Peters, M., H\r{a}berg, S., Xu, Z., van Meurs, J., Gaunt, T.,
  Kerkhof, M., Corpeleijn, E., Feinberg, A., Eng, C., Baccarelli, A., Neelon,
  S.~B., Bradman, A., Merid, S., Bergstr\"om, A., Herceg, Z., Hernandez-Vargas,
  H., Brunekreef, B., Pinart, M., Heude, B., Ewart, S., Yao, J., Lemonnier, N.,
  Franco, O., Wu, M., Hofman, A., McArdle, W., {Van~der~Vlies}, P., Falahi, F.,
  Gillman, M., Barcellos, L., Kumar, A., Wickman, M., Guerra, S., Charles,
  M.-A., Holloway, J., Auffray, C., Tiemeier, H., Smith, G., Postma, D.,
  Hivert, M.-F., Eskenazi, B., Vrijheid, M., Arshad, H., Ant\'o, J., Dehghan,
  A., Karmaus, W., Annesi-Maesano, I., Sunyer, J., Ghantous, A., Pershagen, G.,
  Holland, N., Murphy, S., DeMeo, D., Burchard, E., Ladd-Acosta, C., Snieder,
  H., Nystad, W., Koppelman, G., Relton, C., Jaddoe, V., Wilcox, A., Mel\'en,
  E., and London, S. (2016).
\newblock DNA methylation in newborns and maternal smoking in pregnancy:
  genome-wide consortium meta-analysis.
\newblock {\em The American Journal of Human Genetics\/}, {\bf 98\/}, 680--696.

\bibitem[{Maechler(2015)}]{Maechler13}
Maechler, M. (2015).
\newblock {\em diptest: Hartigan's Dip Test Statistic for Unimodality -
  Corrected\/}.
\newblock R package version 0.75-7.
\urlprefix\url{http://CRAN.R-project.org/package=diptest}

\bibitem[{McLachlan and Peel(2000)McLachlan \& Peel}]{McLachlan00}
McLachlan, G., and Peel, D. (2000).
\newblock {\em Finite Mixture Models\/}.
\newblock John Wiley \& Sons, United States of America.

\bibitem[{McQuillan \emph{et~al.}(2014)McQuillan, Mazeh, \&
  Aigrain}]{mcquillan14}
McQuillan, A., Mazeh, T., and Aigrain, S. (2014).
\newblock Rotation periods of 34,030 Kepler main-sequence stars: the full
  autocorrelation sample.
\newblock {\em The Astrophysical Journal Supplement Series\/}, {\bf 211\/}, 24.

\bibitem[{Minnotte \emph{et~al.}(1998)Minnotte, Marchette, \&
  Wegman}]{Minetal98}
Minnotte, M.~C., Marchette, D.~J., and Wegman, E.~J. (1998).
\newblock The bumpy road to the mode forest.
\newblock {\em Journal of Computational and Graphical Statistics\/}, {\bf 7\/},
  239--251.

\bibitem[{Minnotte and Scott(1993)Minnotte \& Scott}]{MinSco93}
Minnotte, M.~C., and Scott, D.~W. (1993).
\newblock The mode tree: A tool for visualization of nonparametric density
  features.
\newblock {\em Journal of Computational and Graphical Statistics\/}, {\bf 2\/},
  51--68.

\bibitem[{M\"uller and Sawitzki(1991)M\"uller \& Sawitzki}]{MulSaw91}
M\"uller, D.~W., and Sawitzki, G. (1991).
\newblock Excess mass estimates and tests for multimodality.
\newblock {\em Journal of the American Statistical Association\/}, {\bf 86\/},
  738--746.

\bibitem[{Oliveira \emph{et~al.}(2014{\natexlab{a}})Oliveira, Crujeiras, \&
  Rodr{\'\i}guez-Casal}]{oliveira14c}
Oliveira, M., Crujeiras, R.~M., and Rodr{\'\i}guez-Casal, A.
  (2014{\natexlab{a}}).
\newblock CircSiZer: an exploratory tool for circular data.
\newblock {\em Environmental and ecological statistics\/}, {\bf 21\/},
  143--159.

\bibitem[{Oliveira \emph{et~al.}(2014{\natexlab{b}})Oliveira, Crujeiras, \&
  Rodr{\'i}guez-Casal}]{Oliveira14b}
Oliveira, M., Crujeiras, R.~M., and Rodr{\'i}guez-Casal, A.
  (2014{\natexlab{b}}).
\newblock {NPCirc}: An {R} Package for Nonparametric Circular Methods.
\newblock {\em Journal of Statistical Software\/}, {\bf 61\/}, 1--26.
\urlprefix\url{http://www.jstatsoft.org/v61/i09/}

\bibitem[{Parzen(1962)}]{Parzen62}
Parzen, E. (1962).
\newblock On estimation of a probability density function and mode.
\newblock {\em The annals of mathematical statistics\/}, {\bf 33\/},
  1065--1076.

\bibitem[{Poncet(2012)}]{Poncet12}
Poncet, P. (2012).
\newblock {\em modeest: Mode Estimation\/}.
\newblock R package version 2.1.
\urlprefix\url{https://CRAN.R-project.org/package=modeest}

\bibitem[{Postman \emph{et~al.}(1986)Postman, Huchra, \& Geller}]{Postman86}
Postman, M., Huchra, J., and Geller, M. (1986).
\newblock Probes of large-scale structure in the Corona Borealis region.
\newblock {\em The Astronomical Journal\/}, {\bf 92\/}, 1238--1247.

\bibitem[{{R Core Team}(2018)}]{R18}
{R Core Team} (2018).
\newblock {\em R: A Language and Environment for Statistical Computing\/}.
\newblock R Foundation for Statistical Computing, Vienna, Austria.
\urlprefix\url{https://www.R-project.org/}

\bibitem[{Richardson and Green(1997)Richardson \& Green}]{Richardson97}
Richardson, S., and Green, P.~J. (1997).
\newblock On Bayesian analysis of mixtures with an unknown number of components
  (with discussion).
\newblock {\em Journal of the Royal Statistical Society. Series B\/}, {\bf
  59\/}, 731--792.

\bibitem[{Roeder(1990)}]{Roeder90}
Roeder, K. (1990).
\newblock Density estimation with confidence sets exemplified by superclusters
  and voids in the galaxies.
\newblock {\em Journal of the American Statistical Association\/}, {\bf 85\/},
  617--624.

\bibitem[{Rosenblatt(1956)}]{Rosenblatt56}
Rosenblatt, M. (1956).
\newblock Remarks on some nonparametric estimates of a density function.
\newblock {\em The Annals of Mathematical Statistics\/}, {\bf 27\/}, 832--837.

\bibitem[{Rufibach and Walther(2010)Rufibach \& Walther}]{rufibach2010}
Rufibach, K., and Walther, G. (2010).
\newblock The block criterion for multiscale inference about a density, with
  applications to other multiscale problems.
\newblock {\em Journal of Computational and Graphical Statistics\/}, {\bf
  19\/}, 175--190.

\bibitem[{Rufibach and Walther(2015)Rufibach \& Walther}]{Rufibach15}
Rufibach, K., and Walther, G. (2015).
\newblock {\em modehunt: Multiscale Analysis for Density Functions\/}.
\newblock R package version 1.0.7.
\urlprefix\url{https://CRAN.R-project.org/package=modehunt}

\bibitem[{Salgado-Ugarte \emph{et~al.}(1998)Salgado-Ugarte, Shimizu, \&
  Taniuchi}]{Salgado98}
Salgado-Ugarte, I.~H., Shimizu, M., and Taniuchi, T. (1998).
\newblock Nonparametric assessment of multimodality for univariate data.
\newblock {\em Stata Technical Bulletin\/}, {\bf 7\/}, 27--35.

\bibitem[{Santoro \emph{et~al.}(2011)Santoro, Beer, Cartus, Schmullius,
  Shvidenko, McCallum, Wegm{\"u}ller, \& Wiesmann}]{santoro11}
Santoro, M., Beer, C., Cartus, O., Schmullius, C., Shvidenko, A., McCallum, I.,
  Wegm{\"u}ller, U., and Wiesmann, A. (2011).
\newblock Retrieval of growing stock volume in boreal forest using
  hyper-temporal series of Envisat ASAR ScanSAR backscatter measurements.
\newblock {\em Remote Sensing of Environment\/}, {\bf 115\/}, 490--507.

\bibitem[{Scott(2015)}]{Scott15}
Scott, D.~W. (2015).
\newblock {\em Multivariate Density Estimation: Theory, Practice, and
  Visualization\/}.
\newblock John Wiley \& Sons, Hoboken, New Jersey.

\bibitem[{Silverman(1981)}]{Silverman81}
Silverman, B.~W. (1981).
\newblock Using kernel density estimates to investigate multimodality.
\newblock {\em Journal of the Royal Statistical Society. Series B\/}, {\bf
  43\/}, 97--99.

\bibitem[{Wand and Jones(1995)Wand \& Jones}]{wandjones}
Wand, M.~P., and Jones, M.~C. (1995).
\newblock {\em Kernel Smoothing\/}.
\newblock Chapman and Hall, Great Britain.

\bibitem[{Watson(1961)}]{Watson61}
Watson, G.~S. (1961).
\newblock Goodness--of--fit tests on a circle.
\newblock {\em Biometrika\/}, {\bf 48\/}, 109--114.

\bibitem[{Weisberg(2005)}]{Weisberg05}
Weisberg, S. (2005).
\newblock {\em Applied Linear Regression\/}.
\newblock John Wiley \& Sons, New York.

\end{thebibliography}

\end{document}